\documentclass[%
 reprint,
superscriptaddress,
 amsmath,amssymb,
 aps,
 prx,
]{revtex4-2}

\usepackage{graphicx}
\usepackage{dcolumn}
\usepackage{bm}
\usepackage{hyperref}
\usepackage{xcolor}
\usepackage{subdepth} 

\usepackage{mathtools}
\usepackage[capitalize]{cleveref}
\crefname{section}{Sec.}{Secs.}
\Crefname{section}{Sec.}{Secs.}
\crefname{appendix}{App.}{Apps.}
\Crefname{appendix}{Appendix}{Apps.}

\newcommand{\norm}[1]{\left\lVert#1\right\rVert}
\newcommand{\LD}[1]{\norm{(i\gamma\tilde{\gamma})_{#1}}}
\newcommand{\LF}[1]{\norm{\gamma_{#1}}}
\newcommand{\embed}[1]{\mathfrak{i}_{#1}}

\DeclareMathOperator{\Tr}{Tr}
\newcommand{\comp}[1]{\bar{#1}}

\newcommand{\stateO}{\mathcal{O}}
\newcommand{\stateE}{\mathcal{E}}

\usepackage[normalem]{ulem}

\begin{document}

\title{Quantifying robustness and locality of Majorana bound states in interacting systems}

\author{William Samuelson}
\email{william.samuelson@ftf.lth.se}
\affiliation{Division of Solid State Physics and NanoLund, Lund University, Lund, Sweden}

\author{Juan Daniel Torres Luna}
\affiliation{QuTech and Kavli Institute of Nanoscience, Delft University of Technology, Delft, 2600 GA, The Netherlands}

\author{Sebastian Miles}
\affiliation{QuTech and Kavli Institute of Nanoscience, Delft University of Technology, Delft, 2600 GA, The Netherlands}

\author{A. Mert Bozkurt}
\affiliation{QuTech and Kavli Institute of Nanoscience, Delft University of Technology, Delft, 2600 GA, The Netherlands}

\author{Martin Leijnse}
\affiliation{Division of Solid State Physics and NanoLund, Lund University, Lund, Sweden}

\author{Michael Wimmer}
\affiliation{QuTech and Kavli Institute of Nanoscience, Delft University of Technology, Delft, 2600 GA, The Netherlands}

\author{Viktor Svensson}
\email{viktor.svensson@fys.uio.no}
\affiliation{Department of Physics, University of Oslo, Oslo, Norway}

\date{\today}

\begin{abstract}
Protecting qubits from perturbations is a central challenge in quantum computing. 
Topological superconductors with separated Majorana bound states (MBSs) provide a strong form of protection that only depends on the locality of perturbations. While the link between MBS separation, robust degeneracy, and protected braiding is well understood in non-interacting systems, recent experimental progress in short quantum-dot-based Kitaev chains highlights the need to establish these connections rigorously for interacting systems. We do this by defining MBSs from many-body ground states and show how their locality constrains their coupling to an environment. This, in turn, quantifies the protection of the energy degeneracy and the feasibility of non-abelian braiding.
\end{abstract}

\maketitle

\section{\label{sec:introduction}Introduction}
Robust degeneracies in quantum systems are of fundamental importance and offer potential for quantum technologies.
Topological protection provides a particularly strong form of robustness, where the degeneracy is insensitive to local perturbations.
Topological superconductors realize this protection as a result of separated Majorana bound states (MBSs): quasiparticles with non-abelian exchange statistics and potential for fault-tolerant quantum computation~\cite{kitaevUnpairedMajoranaFermions2001}; see Refs.~\cite{wilczekMajoranaReturns2009,leijnseIntroductionTopologicalSuperconductivity2012,aliceaNewDirectionsPursuit2012,aguadoMajoranaQuasiparticlesCondensed2017,beenakkerSearchMajoranaFermions2013} for comprehensive overviews.
Although conclusive experimental evidence for topological MBSs is still lacking, the field has seen substantial theoretical and experimental development. On the theoretical side, progress includes increasingly realistic platform proposals~\cite{lutchynMajoranaFermionsTopological2010,oregHelicalLiquidsMajorana2010,nadj-pergeProposalRealizingMajorana2013,hellTwoDimensionalPlatformNetworks2017,pientkaTopologicalSuperconductivityPlanar2017,sauRealizingRobustPractical2012,fulgaAdaptiveTuningMajorana2013,leijnseParityQubitsPoor2012,liuTunableSuperconductingCoupling2022}, as well as braiding protocols, readout schemes and network architectures~\cite{aliceaNonAbelianStatisticsTopological2011,aasenMilestonesMajoranaBasedQuantum2016,tsintzisMajoranaQubitsNonAbelian2024,karzigScalableDesignsQuasiparticlepoisoningprotected2017}. 

In non-interacting models, the spatial separation of MBSs implies both protected braiding protocols and robustness of the ground-state degeneracy against local perturbations.
Extensions of these concepts to interacting systems include the classification of topological phases~\cite{fidkowskiEffectsInteractionsTopological2010,fidkowskiTopologicalPhasesFermions2011}, numerical and perturbative studies of phase diagrams~\cite{stoudenmireInteractionEffectsTopological2011,hasslerStronglyInteractingMajorana2012,gergsTopologicalOrderKitaev2016,aksenovStrongCoulombInteractions2020,chepigaTopologicalQuantumCritical2023},
and exact solutions of interacting models at fine-tuned points~\cite{katsuraExactGroundStates2015,miaoExactSolutionInteracting2017}. Furthermore, the many-body structure of the MBSs themselves has been studied~\cite{goldsteinExactZeroModes2012,wrightLocalizedManyParticleMajorana2013,obrienManyparticleMajoranaBound2015,obrienManybodyInterpretationMajorana2015,kellsManybodyMajoranaOperators2015,kellsMultiparticleContentMajorana2015,mcginleyRobustnessMajoranaEdge2017,kellsLocalizationEnhancedDegraded2018,bozkurtInteractioninducedStrongZero2025,svenssonQuantumDotBased2024} and they may contain a significant amount of non-local and many-body contributions. It has remained unclear how to quantify the locality of such operators in a meaningful way, and when to expect interacting systems to exhibit non-abelian braiding.
In light of recent experimental progress in short quantum-dot-based Kitaev chains~\cite{dvirRealizationMinimalKitaev2023,tenhaafTwositeKitaevChain2024,zatelliRobustPoorMans2024,bordinEnhancedMajoranaStability2025,tenhaafObservationEdgeBulk2025,bordinProbingMajoranaLocalization2025}, where the MBS separation must be fine-tuned and interactions are strong, it is increasingly important to understand this quantitatively.

\begin{figure}[t!]
    \centering
\includegraphics[width=1\linewidth]{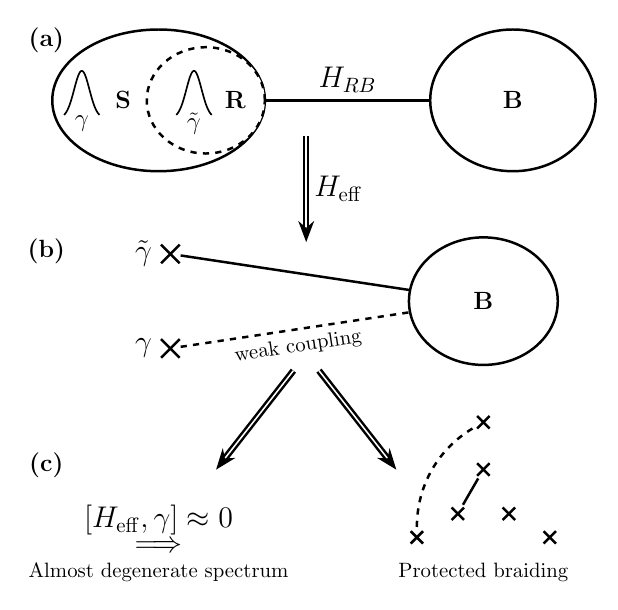}
    \caption{We show that the separation of MBSs implies energy protection and braiding, using methods applicable in interacting systems. (a). We consider a system $S$, described by two MBSs. It interacts with an environment $B$, with a coupling acting in a subregion $R$. 
    (b). The couplings to the environment are bounded by the localization of the MBSs, quantified by the partial trace.
    (c). A weakly coupled MBS almost commutes with the Hamiltonian, implying stability of the energy spectrum. The localization of the MBSs also determines the feasibility of non-abelian braiding.
    }
    \label{fig:system}
\end{figure}

The structure and main logic of our approach are summarized in \cref{fig:system}.
In \cref{sec:interacting_majoranas_definition}, we define ground-state MBSs from a pair of low-energy states with opposite fermion parity. In \cref{sec:coupling_bounds}, we derive bounds on the effective coupling between such MBSs and an environment. These bounds have direct consequences for the stability of the spectrum, which we show in \cref{sec:energy_splitting_bounds}. We obtain physically meaningful measures of MBS locality from partial traces, and in \cref{sec:gauge}, we show how it connects to the Majorana polarization used in the literature before.
Considering the interacting Kitaev chain in \cref{sec:example}, we analytically describe the structure of the MBSs and demonstrate our bounds on energy splittings in the case of disorder and when coupling to an external dot.
In \cref{sec:example_abs_to_ysr}, we study an experimentally realistic system, a quantum-dot-based Kitaev chain, linking our quality measures to experimental measurements. In \cref{sec:braiding}, we couple multiple systems together and show how our measures have consequences for the success of coupling-based braiding protocols. 
The appendices contain rigorous derivations and extended technical discussions.

\section{MBS\MakeLowercase{s} in interacting systems}\label{sec:interacting_majoranas_definition}
In non-interacting systems, single-particle MBSs are obtained by diagonalizing the Bogoliubov-de-Gennes Hamiltonian. These operators act on the full Hilbert space. However, in adiabatic braiding protocols, only the low-energy ground-state sector is relevant. Motivated by this fact, we consider a fermionic system $S$ with two low-energy states of opposite fermion parity, $|e\rangle$ and $|o\rangle$, separated from all excited states by an energy gap. 

As an operator basis for this space, we define
\begin{subequations}\label{eq:ground_state_majoranas_definition}
\begin{align}
 \gamma &= |e\rangle\langle o| + |o\rangle\langle e|, \\
    \tilde{\gamma} &= i (|e\rangle\langle o| - |o\rangle\langle e|),\\
    i\gamma\tilde{\gamma} &= |e\rangle \langle e| - |o\rangle \langle o|,
\end{align}
\end{subequations}
which are analogous to the Pauli operators. They are hermitian and square to identity in the space of ground states
\begin{align}
    \gamma^2 = \tilde{\gamma}^2 = |e\rangle \langle e| + |o\rangle \langle o| \equiv Q.
\end{align}
A crucial difference with spin operators is that $\gamma$ and $\tilde\gamma$ are fermionic. We refer to them as ground-state MBSs, and their product as their parity. These operators describe the low-energy physics of $S$, and can be defined in both interacting and non-interacting systems.

The ground-state MBSs are 'weak' in the sense that they only map between the opposite parity states in the ground-state sector \cite{aliceaTopologicalPhasesParafermions2016,wadaCoexistenceStrongWeak2021}. This is in contrast to 'strong' MBSs, which map between the parity sectors in the full Hilbert space~\cite{fendleyStrongZeroModes2016, kellsManybodyMajoranaOperators2015,kellsMultiparticleContentMajorana2015,kellsLocalizationEnhancedDegraded2018,behrendsTenfoldWayManybody2019,chepigaTopologicalQuantumCritical2023,bozkurtInteractioninducedStrongZero2025}. Whether or not a strong MBS exists is irrelevant here---our focus is solely on the ground-state sector.

Unlike single-particle MBSs, $\gamma$ and $\tilde\gamma$ are non-local and generally act non-trivially on the entire system. Nevertheless, when coupling to them in a region $R$, only the local part plays a role.
This is captured by the fermionic partial trace, which defines the reduced MBSs in region $R$ as
\begin{align}\label{eq:partial_trace}
    \gamma_R = \mathrm{Tr}_{\overline{R}}[\gamma], 
\end{align}
where $\comp{R}$ is the complement of the region $R$. For fermionic systems, partial traces and tensor products need to be slightly modified to ensure consistency with the anti-commutation relations, and we follow Ref.~\cite{szalayFermionicSystemsQuantum2021} closely. These technicalities are not important for the conceptual understanding of our results, so we have saved them for the proofs in the Appendix. 

Due to fermionic superselection, the relative phase between $|e\rangle$ and $|o\rangle$ is an unphysical gauge freedom. A rotation of this phase corresponds to mixing the two ground-state MBSs, and it is important to choose the phase to maximize the separation of $\gamma$ and $\tilde\gamma$ in the coupling region $R$. This leads to a many-body generalization of the Majorana polarization, see \cref{sec:gauge}.

\section{Bounding couplings by locality}\label{sec:coupling_bounds}
To see how the localization of the ground-state MBSs implies protection from perturbations, we couple the system $S$ to an environment $B$. The full Hamiltonian reads 
\begin{equation}\label{eq:coupled_ham}
    H = H_S + H_{RB} + H_B,
\end{equation}
where $H_{RB}$ is assumed to act only in a subregion $R$ of $S$ and $H_B$ is arbitrary. The coupling can be split into odd and even terms,
\begin{equation}
    H_{RB} = H_{RB}^e + H_{RB}^o.
\end{equation}
$H_{RB}^e$ has terms with even fermion parity in both $R$ and $B$, such as local chemical potential shifts within $R$ or Coulomb interaction between $R$ and $B$, while $H_{RB}^o$ collects terms with odd parity in both $R$ and $B$, such as single-particle hopping between them.

Projecting out the excited states in $S$, we can write an effective Hamiltonian in terms of the operators in \cref{eq:ground_state_majoranas_definition}. It takes the form 
\begin{equation}\label{eq:effective_ham}
    H_\text{eff} = (\varepsilon + G) i\gamma \tilde{\gamma}/2 + \gamma F/2 + \tilde{\gamma} \tilde{F}/2 + Q\hat{H}_B ,
\end{equation}
where $F, \tilde{F}$, and $G$ are operators acting in the environment \footnote{To simplify notation in the main text, we do not write explicitly on which space operators act. The convention we use is that it acts on the smallest space possible allowed in the context. When two operators in different spaces are multiplied or added, they are implicitly extended to a joint space by a fermionic tensor product.}, $\varepsilon$ is the energy difference between the low-energy states of $H_S$, and $\hat{H}_B$ is a renormalized Hamiltonian for the environment. 
To leading order in the system-bath coupling, the effective coupling Hamiltonian is $QH_{RB}Q$. This is the projection of a local operator, which we will exploit to bound the coupling to the ground-state MBSs. Higher-order terms are generally more non-local~\cite{bravyiSchriefferWolff2011} and would require additional analytical tools to treat systematically. In our derivation, we assume weak coupling and neglect those contributions. In the numerical examples, we consider larger coupling strengths to see possible violations of our results.

Using the leading-order projected Hamiltonian, the effective operators can be expressed as
\begin{subequations}\label{eq:effective_ops}
\begin{align}
    F &= \Tr_S{\left[\gamma H_{RB}^o\right]} = \Tr_R{\left[\gamma_R H_{RB}^o\right]}, \\
    \tilde{F} &= \Tr_S{\left[\tilde{\gamma} H_{RB}^o\right]} = \Tr_R{\left[\tilde{\gamma}_R H_{RB}^o\right]}, \\
    G &= \Tr_S{\left[i\gamma\tilde{\gamma} H_{RB}^e\right]} = \Tr_R{\left[(i\gamma\tilde{\gamma})_R H_{RB}^e\right]},
\end{align}
\end{subequations}
where $\gamma_R$ is the reduced ground-state MBS defined in \cref{eq:partial_trace}.
If the norms of the reduced MBSs are small, the effective operators will be small.
In \cref{app:general_effective_operator_bounds}, we prove that 
\begin{subequations}\label{eq:effective_operators_bound}
\begin{align}
    \norm{F}_p &\leq \norm{H_{RB}^o}_p \norm{\gamma_R}_q, \\
    \lVert{\tilde{F}\rVert}_p &\leq \norm{H_{RB}^o}_p \norm{\tilde{\gamma}_R}_q, \\
    \norm{G}_p &\leq \norm{H_{RB}^e}_p \norm{(i\gamma\tilde{\gamma})_R}_q, 
\end{align}
\end{subequations}
where $1/q + 1/p = 1$, and we use the Schatten norm defined by
\begin{equation}
    \norm{A}_p = \Tr \left[\sqrt{A^\dagger A}^p\right]^{1/p}.
\end{equation}
These bounds are sharp. For example, for $q=2$ (the Frobenius norm), the bound for $F$ is saturated if
\begin{equation}\label{eq:optimal_coupling}
    H_{RB}^o = \gamma_R h_B,
\end{equation}
for any operator $h_B$ in the environment. 
 
These bounds constrain how much an environment can couple to a pair of MBSs and their parity.
We use these to derive bounds on the energy spectrum in \cref{sec:energy_splitting_bounds} and discuss consequences for braiding in \cref{sec:braiding}.

\section{Energy splitting bound}\label{sec:energy_splitting_bounds}
A useful property of MBSs is that the energy spectrum is protected if they commute with a perturbation. We derive an inequality connecting the protection to the localization of the ground-state MBSs, see \cref{app:energy_splitting_bounds} for all the details. While the case of most interest is when $S$ is degenerate ($\varepsilon=0$), we formulate the more general result with non-zero initial splitting $\varepsilon$. These bounds only hold in the low-energy subspace---they can be violated if the coupling is comparable to the excitation gap in $S$.

Let $|\stateO\rangle$ and $|\stateE\rangle$ be eigenstates of $H_\text{eff}$, with odd and even parity, respectively. Their energy difference is 
\begin{equation}\label{eq:energy_difference_definition}
    \delta E =  \langle \stateE | H_\text{eff} |\stateE \rangle - \langle \stateO | H_\text{eff} | \stateO \rangle = \frac{\langle \stateO | [\gamma, H_\text{eff}]  | \stateE \rangle}{\langle \stateO | \gamma | \stateE \rangle}.
\end{equation}
The commutator of the effective Hamiltonian and the ground-state MBS $\gamma$ is 
\begin{equation}\label{eq:effective_commutator_interacting}
    [\gamma, H_\text{eff}] =  \varepsilon i\tilde{\gamma}+ i\tilde{\gamma}G  +  F.
\end{equation}
Using \cref{eq:effective_operators_bound,eq:energy_difference_definition,eq:effective_commutator_interacting}, we obtain 
\begin{multline} \label{eq:non_perturbative_bound}
    |\delta E\langle \stateO | \gamma | \stateE\rangle - \varepsilon \langle\stateO | i\tilde{\gamma} | \stateE\rangle | \leq  \LD{R}_q \norm{H_{RB}^e}_p \\
    + \LF{R}_q \norm{H_{RB}^o}_p \norm{(|\stateE\rangle\langle\stateO|)_B}_q.
\end{multline}
The operator $|\stateE\rangle\langle\stateO|$ can be thought of as a fermion that switches between the two states, and $\norm{(|\stateE\rangle\langle\stateO|)_B}$ quantifies how much of that fermion is contained in the environment.

We can specify additional bounds on the three quantities $|\langle\stateO | \gamma | \stateE\rangle|, |\langle\stateO | i\tilde{\gamma} | \stateE\rangle|$ and $\norm{(|\stateE\rangle\langle\stateO|)_B}_q$. 
In \cref{app:energy_splitting_bounds}, we derive several bounds applicable in different regimes set by the ratios of the energy scales $\varepsilon$, $\norm{H_{RB}}$, and the level spacings in $B$.

Here, we assume that $\norm{H_{RB}}$ may be larger than $\varepsilon$ and any level spacings in $B$, but still smaller than the excitation gap in $S$. If there are $d_B$ states in $B$, we find the result:
For any eigenstate of $H_\text{eff}$, there exists another eigenstate with opposite parity such that their energy difference is bounded by
\begin{equation}\label{eq:perturbative_bound}
    |\delta E| \leq \sqrt{d_B}(\LD{R}_q \norm{H_{RB}^e}_p + \LF{R}_q \norm{H_{RB}^o}_p+ |\varepsilon|).
\end{equation}

There is a straightforward generalization to the case where the coupling is written as a sum of terms 
\begin{equation}
    H_c = \sum_n H_{R_nB}
\end{equation}
acting in different regions $R_n$, see \cref{app:sumofterms}.
While the above bound can be used with $R = \bigcup_n R_n$, a stronger bound is obtained by associating each term with a set of effective operators, each satisfying \cref{eq:effective_operators_bound}. The commutator then becomes  
\begin{equation}\label{eq:effective_commutator_interacting_many_terms}
    [\gamma, H_\text{eff}] = \varepsilon i\tilde{\gamma} + \sum_n i\tilde{\gamma}G_n  +  F_n,
\end{equation}
and the bound on the energy splitting is 
\begin{align} \label{eq:non_perturbative_bound_many_terms}
    |\delta E\langle \stateO | \gamma | \stateE\rangle - \varepsilon \langle\stateO | i\tilde{\gamma} | \stateE\rangle | &\leq \norm{\sum_n i\tilde{\gamma}G_n  +  F_n}_p \nonumber
    \\ 
    \leq \sum_n \LD{R_n}_q& \norm{H_{R_nB}^e}_p \nonumber \\
     + \LF{R_n}_q& \norm{H_{R_nB}^o}_p \norm{(|\stateE\rangle\langle\stateO|)_B}_q.
\end{align}
We can simplify this to
\begin{equation} \label{eq:simple_general_energy_bound}
    |\delta E| \leq \sqrt{d_B}\left(Q_e h_e + Q_o h_o
     + |\varepsilon|\right),
\end{equation}
where 
\begin{subequations}
\begin{align}
    Q_o &= \sqrt{\sum_{n} \norm{\gamma_{R_n}}_q^2},\\
    Q_e &= \sqrt{\sum_{n}\norm{(i\gamma \tilde{\gamma})_{R_n}}_q^2}, \\
    h_o &= \sqrt{\sum_n \norm{H_{R_{n}B}^o}_p^2}, \\
    h_e &= \sqrt{\sum_n \norm{H_{R_{n}B}^e}_p^2}.
\end{align}
\end{subequations}
The MBS quality factors $Q_o$ and $Q_e$ quantify the protection against odd and even couplings and depend only on the subregions in which they act, while $h_o$ and $h_e$ measure the corresponding coupling strengths.


\section{Quality measures, gauge choice and Majorana polarization}\label{sec:gauge}
Here, we consider how the measures depend on the relative phase between $|e\rangle$ and $|o\rangle$, which is a gauge freedom due to superselection. The gauge rotates the ground-state MBSs into each other, but their parity, $i\gamma\tilde\gamma$, is invariant. The local parity can be written in terms of the reduced density matrices of the odd ($\rho_R^o$) and even ($\rho_R^e$) ground states as
\begin{equation}\label{eq:LD}
    \norm{(i\gamma\tilde\gamma)_R}_q = \norm{\rho_R^e - \rho_R^o}_q,
\end{equation}
and thus measures the local distinguishability of the two ground states in $R$, as discussed in Ref.~\cite{svenssonQuantumDotBased2024}. 

The ground-state MBSs, the couplings $F$ and $\tilde{F}$, and $Q_o$ are gauge-dependent. Some gauges may give better bounds than others. To get the strongest bound in \cref{eq:simple_general_energy_bound}, we minimize $Q_o$.
For $q=2$, the minimization admits an analytical solution.
Defining the ground-state fermion 
\begin{equation}    
c = (\gamma + i\tilde{\gamma})/\sqrt{2}, 
\end{equation}
we find the minimum to be
\begin{equation}
\begin{aligned}
    \min_{\text{gauges}} \sum_n\norm{\gamma_{R_n}}_2^2 
    = \sum_n\Tr[c^\dagger_{R_n} c_{R_n}] - \Bigl\lvert\sum_n \Tr[c_{R_n} c_{R_n}]\Bigr\rvert,
\end{aligned}
\end{equation}
see \cref{app:optim_gauge} for details.
The first term can be interpreted as a fermion density, and the second as a Majorana density. We will see that the ratio of these,
\begin{equation}\label{eq:many_body_normalized_majorana_polarization}
    M = \frac{\Bigl\lvert\sum_n \Tr[c_{R_n} c_{R_n}]\Bigr\rvert}{\sum_n\Tr[c^\dagger_{R_n} c_{R_n}]},
\end{equation}
is a generalization of the Majorana polarization, a commonly used measure of MBS quality. This quantity depends on the two ground states and on the set of regions $\{R_n\}$ the coupling acts on.

The maximum value of $M$ is 1, in which case $Q_o=0$ and one can only couple to a single MBS in this set of subregions. 
The phase of $\Tr[c_{R_n} c_{R_n}]$ determines the optimal gauge for the subregion $R_n$. There is only one gauge parameter, so if it varies, we cannot optimize each term simultaneously. Then, the value of $M$ decreases, implying less protection against simultaneous odd-parity couplings to those regions. As an example, if the environment couples via tunneling to both ends of a symmetric chain, the terms in the numerator cancel, giving $M=0$.

If $H_S$ is real, the optimal gauge will either be when the eigenstates are real ($\gamma$ is the smallest) or purely imaginary ($\tilde{\gamma}$ is the smallest).

\subsection{Comparison with existing quality measures}
\subsubsection{Interacting systems}
Here, we consider a special case where the general expression in \cref{eq:many_body_normalized_majorana_polarization} reduces to an expression often used to quantify MBS quality in interacting systems. Let $f_j$ be a standard fermion and define the Majorana basis operators
\begin{subequations}\label{eq:sp_majorana_basis}
    \begin{align}
        \Gamma_j^+ &= f^\dagger_j + f_j, \\
        \Gamma_j^- &= i (f^\dagger_j - f_j).
    \end{align}
\end{subequations}
Consider an odd coupling which can be written as a sum over single-mode subregions $R_j$ and let $R = \cup_j R_j$. 
The coupling takes the form
\begin{equation}
    H_c^o = \sum_{j \in R} t_j f_j^\dagger F_j + \text{h.c.},
\end{equation}
where $F_j$ is any odd operator in the environment. This corresponds to tunneling of $f$-particles in and out of the system. For single-mode subregions, the reduced MBSs are fully defined by the matrix elements $\langle o|\Gamma_j^s|e\rangle$, and for this special case, \cref{eq:many_body_normalized_majorana_polarization} becomes 
\begin{equation}\label{eq:majorana_polarization_ground_state_matrix_elements}
    M = \frac{\left| \sum_{s,j \in R} \langle o|\Gamma_j^s| e\rangle^2 \right|}{\sum_{s,j \in R} |\langle o|\Gamma_j^s|e\rangle|^2},
\end{equation}
which has been used before \cite{tsintzisMajoranaQubitsNonAbelian2024,aksenovStrongCoulombInteractions2020, tsintzisCreatingDetectingPoor2022, soutoProbingMajoranaLocalization2023,samuelsonMinimalQuantumDot2024,torreslunaFluxtunableKitaevChain2024}. This puts the use of \cref{eq:majorana_polarization_ground_state_matrix_elements} on a firm footing as it bounds the energy splitting coming from the most physically relevant odd couplings. The only case where it does not apply is when the odd coupling has many-body terms in $S$. The single-mode subregions do not have to be purely spatial: in spinful systems, they refer to a single spin at one site. 


\subsubsection{Non-interacting systems}\label{sec:non-interacting}
If the system is non-interacting, 
one can derive bounds without projecting to the ground-state sector (see \cref{app:non-interacting_bounds}). 
These bounds involve the single-particle MBSs $\Gamma$ and $\tilde\Gamma$ obtained by diagonalizing the BdG-Hamiltonian. Note that even in non-interacting systems, these are very different from the ground-state MBSs $\gamma$ and $\tilde{\gamma}$, see \cref{app:manybody_desc} for an extended discussion.

The derivation is analogous to the interacting case, yielding similar bounds for the couplings and energy splitting. In this case, we will use a single region $R$, because the single-particle structure will allow us to split the region into individual fermionic modes anyway. The quality measures are then $Q_o = \norm{\Gamma_R}_2$, which bounds odd couplings, and $Q_e = \lVert(i\Gamma\tilde{\Gamma})_R\rVert_2$, which bounds the even couplings. There is an analogous gauge freedom here, corresponding to a Bogoliubov rotation of the two single-particle MBSs. Again, $Q_o$ is gauge-dependent, but $Q_e$ is not.
By optimizing the gauge, we find
\begin{equation}
    \min_\mathrm{gauges} \norm{\Gamma_R}_2^2 = \Tr[C^\dagger_{R} C_{R}](1 - M),
\end{equation}
where $C = (\Gamma + i \tilde\Gamma)/\sqrt{2}$ and 
\begin{equation}\label{eq:majorana_polarization_sp_fermion}
    M = \frac{\bigl\lvert\Tr[C_{R} C_{R}]\bigr\rvert}{\Tr[C^\dagger_{R} C_{R}]}
\end{equation}
is the Majorana polarization from \cref{eq:many_body_normalized_majorana_polarization} when using a single region $R$ and the single-particle fermion $C$. In contrast to the interacting case, the single-particle structure of the MBSs can be used to simplify the local parity as
\begin{equation}
\norm{(i\Gamma\tilde{\Gamma})_R}_2^2 = \frac{\Tr[C^\dagger_{R} C_{R}]^2}{d_Rd_{\comp{R}}^2}\left(1 - M^2\right).
\end{equation}
So in the non-interacting case, the Majorana polarization can be used to bound the energy splitting for both odd and even couplings.

We can rewrite \cref{eq:majorana_polarization_sp_fermion} in terms of the single-particle wavefunction. Writing the fermion as
\begin{equation}
    C = \sum_j u_j f_j + v_j f_j^\dagger,
\end{equation}
and letting $\mathbf{u}_R$ and $\mathbf{v}_R$ be vectors containing the coefficients belonging to operators acting on the subregion $R$, the normalized Majorana polarization is
\begin{equation}\label{eq:majorana_polarization_in_terms_of_single_particle_wavefunction}
    M = \frac{2|\mathbf{v}_R^T\mathbf{u}_R|}{\mathbf{u}_R^\dagger\mathbf{u}_R + \mathbf{v}_R^\dagger\mathbf{v}_R}.
\end{equation}
Originally, the complex number $2v_j u_j$ was named Majorana polarization \cite{sedlmayrVisualisingMajoranaBound2015, sedlmayrMajoranaBoundStates2016, benaTestingFormationMajorana2017a}, where it was stressed that the variation of the phase implies lower quality MBSs. The same feature occurs here in the many-body generalization $\Tr[c_R c_R]$, where the phase variation means that the optimal gauge is different in different regions.

\Cref{eq:majorana_polarization_ground_state_matrix_elements,eq:majorana_polarization_in_terms_of_single_particle_wavefunction,eq:majorana_polarization_sp_fermion} are equivalent in non-interacting systems. \Cref{eq:majorana_polarization_ground_state_matrix_elements} can therefore be used to bound the energy splitting in both non-interacting and interacting systems, though for the latter only in a special (but physically relevant) case and only for the odd coupling.

\section{Example I: interacting Kitaev chain}
\label{sec:example}
To illustrate our results, we take the interacting Kitaev chain~\cite{gergsTopologicalOrderKitaev2016,katsuraExactGroundStates2015,miaoExactSolutionInteracting2017,kellsLocalizationEnhancedDegraded2018}
\begin{equation}\label{eq:kitaev_chain}
    H_K = \sum_{j=1}^N \mu_j n_j + \sum_{j=1}^{N-1} (t f_j^\dagger f_{j+1} + \Delta f_j^\dagger f_{j+1}^\dagger + \text{h.c.})+ U n_{j} n_{j+1},
\end{equation}
where $n_j = f_j^\dagger f_j$, $\mu_j$ is the chemical potential at site $j$, $t$ and $\Delta$ are the hopping and pairing amplitudes (which we take to be real), and $U$ is the nearest-neighbor Coulomb interaction.

First, we analyze $H_K$ analytically at a fine-tuned sweet spot where MBSs are perfectly localized on the outermost sites. Second, we study the isolated chain away from the sweet spot and demonstrate the decay of the MBSs away from the edges. Finally, we test the bounds derived in \cref{sec:energy_splitting_bounds} when adding disorder to the chain, and when coupling the chain to a quantum dot. 
We form ground-state MBSs using the real eigenstates. 

\subsection{Ground-state MBSs at the sweet spot}\label{sec:example_sweet_spot}
At the point
\begin{subequations}\label{eq:sweetspot}
\begin{align}
    \Delta &= t + U/2, \\
    \mu_j &= \begin{cases}
        -U,  &1<j<N \\
        -U/2, &j=1,N
    \end{cases},
\end{align}
\end{subequations}
the ground states of $H_K$ are degenerate and take the same form as the ground states in the non-interacting Kitaev chain ($U=0$) at $\mu_j=0, t=\Delta$~\cite{katsuraExactGroundStates2015}. 
In the non-interacting limit, the standard, single-particle MBSs are localized at the edges. The situation is similar for the ground-state MBSs in the interacting case, although with additional subtleties.

As a concrete example, we take $N = 3$. Then, in terms of the single-particle Majorana basis $\Gamma_j^s$ defined in \cref{eq:sp_majorana_basis}, 
we find
\begin{equation}
    \gamma = \Gamma_1^+(I - i\Gamma_1^-\Gamma_2^+)(I - i\Gamma_2^-\Gamma_3^+)/4.
\end{equation}
This operator has support on the full system, but the partial traces to single sites are
\begin{subequations}
\begin{align}
    \gamma_1 &= \Tr_{\{2,3\}} [\gamma] = \Gamma_1^+, \\
    \gamma_2 &= \Tr_{\{1,3\}} [\gamma] = 0, \\
    \gamma_3 &= \Tr_{\{1,2\}} [\gamma] = 0.
\end{align}
\end{subequations}
The other ground-state MBS also has support on the full system, but its partial traces are nonzero only for regions that include the third site. Therefore, one can only couple to $\gamma$ at the first site, and to $\tilde\gamma$ on the third site. This example highlights that the localization of these operators is subtle. For further discussion of these points, see \cref{app:manybody_desc}.

\subsection{Spatial profile of ground-state MBSs}\label{sec:example_isolated_chain}

\begin{figure}[t!]
    \centering
    \includegraphics[width=1\linewidth]{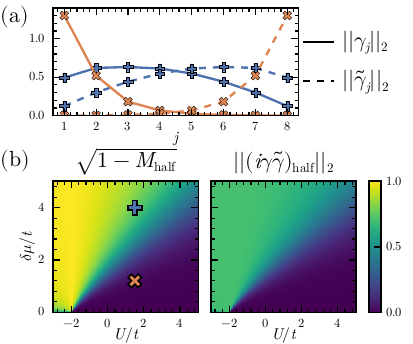}
    \caption{Quality measures and locality of ground-state MBSs for an 8-site interacting Kitaev chain. The spatial profile of the norm of the reduced MBSs is plotted in (a), for the two sets of parameters marked in (b).
    The heatmaps in (b) show the Majorana polarization and the local parity for a single region covering the left half of the system. $\delta\mu$ is a global detuning of the chemical potential away from the interacting sweet spot in \cref{eq:sweetspot}. 
    The quality measures quantify the protection of the system against perturbations acting on half the chain, and give a hint of the phase diagram of this model.}
    \label{fig:wavefunctions_and_heatmaps}
\end{figure}

If the chemical potential is detuned away from the sweet spot, one expects the MBSs to decay away from the edges. To confirm this intuition, we take the subregion to be a single site $j$ and consider the quantities $\lVert\gamma_j\rVert_2$, $\lVert\tilde\gamma_j\rVert_2$. These are plotted in \cref{fig:wavefunctions_and_heatmaps}a, for an 8-site chain, for the parameters marked in \cref{fig:wavefunctions_and_heatmaps}b. Since the MBSs are gauge-dependent, so is their overlap, and it is important to pick the optimal gauge to interpret the overlap as a lack of protection. In this example, the MBSs we plot are the optimal ones, but the smallest MBS (or equivalently, the optimal gauge) switches abruptly in the middle of the chain.

In \cref{fig:wavefunctions_and_heatmaps}b, we take $R$ to be one half of the chain and plot the gauge-invariant quality measures to show the phase diagram of the interacting Kitaev chain around the sweet spot. As $U$ is varied, both $\Delta$ and $\mu$ vary with it according to \cref{eq:sweetspot}. We then add a global detuning $\delta\mu$ of the chemical potential. The exact sweet spot occurs along the whole $x$-axis where $\delta\mu=0$. In the dark region, the chain is robust against arbitrary perturbations and couplings acting on the left half of the system.

\subsection{Energy bounds under disorder}
As a first illustration of the bounds on the energy splitting, we consider the chemical potential to be fluctuating along the chain. There is no environment, or equivalently, $B$ is 1-dimensional. In this example and the next, we take the Hamiltonian of the system as \cref{eq:kitaev_chain} with the parameters
\begin{equation}\label{eq:HS_parameters}
\begin{aligned}
t&=\Delta=U/2, \\
\mu_j &\approx \begin{cases}
-2.73t  \text{ at the edges,} \\
-5.46t  \text{ in the bulk,}
\end{cases}
\end{aligned}
\end{equation}
and $N=8$ sites.
The chemical potential is chosen according to Ref.\,\cite{katsuraExactGroundStates2015} to get degenerate ground states. The MBSs have a small overlap (see the quality measures below) and the gap to excited states is approximately $t/2$.

\begin{figure}[t!]
    \centering
    \includegraphics[width=1\linewidth]{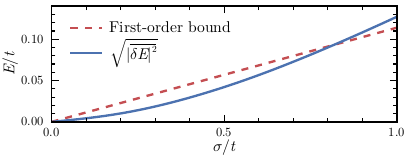}
    \caption{Root-mean-square energy splitting in an 8-site interacting Kitaev chain as a function of the disorder strength $\sigma$. The parameters are given in \cref{eq:HS_parameters}. The bound \cref{eq:disorder_bound} holds rigorously at first order in $\sigma/t$. When the disorder is comparable to the energy gap and higher-order effects are important, the bound is violated.}
    \label{fig:disorder}
\end{figure}

The coupling Hamiltonian is
\begin{equation}
    H_c = \sum_j H_{R_jB} = \sum_{j=1}^N \delta\mu_j n_j,
\end{equation}
where we draw $\delta\mu_j$ from independent normal distributions with variance $\sigma^2$.
To bound the energy splitting, we use \cref{eq:simple_general_energy_bound} with $h_o = 0$ and subregions $R_j = j$, since $H_c$ is a sum of terms acting on each site. The bound is then
\begin{equation}
    |\delta E|^2 \leq Q_e^2 \sum_{j=1}^N |\delta \mu_j|^2,
\end{equation}
where $Q_e^2 = \sum_j \lVert (i\gamma\tilde\gamma)_j \rVert^2_q$. Taking the expectation value and using the independence of $\delta\mu_j$, we obtain a bound on the root-mean-square splitting,
\begin{equation}\label{eq:disorder_bound}
    \sqrt{\overline{|\delta E|^2}} \leq \sqrt{N} Q_e \sigma. 
\end{equation}

In this example, the strongest bound is obtained by choosing $q=\infty$, where $Q_e\approx 0.04$. \Cref{fig:disorder} shows a numerical calculation using 1000 disorder realizations for different values of $\sigma$. The bound holds for small disorder strengths and is violated for strengths comparable to the excitation gap ($\sim t/2$). 

\subsection{Energy bounds when coupled to a quantum dot}\label{sec:example_coupling}
\begin{figure}[t!]
    \centering
    \includegraphics[width=1\linewidth]{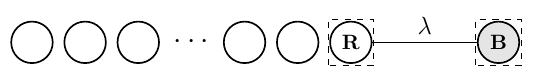}
    \includegraphics[width=1\linewidth]{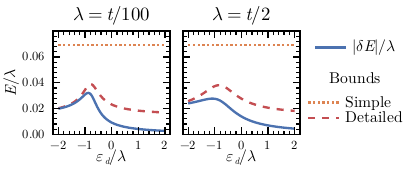}
    \caption{The ground-state energy splitting $\delta E$ (solid) of an (initially degenerate) 8-site interacting Kitaev chain, when coupled weakly ($\lambda=t/100$) and strongly ($\lambda=t/2$) to a quantum dot, varying the dot level $\varepsilon_d$. The dashed and dotted lines show the bounds in \cref{eq:non_perturbative_bound_ex} (Detailed) and \cref{eq:perturbative_bound_ex} (Simple).
    The system Hamiltonian is defined by the parameters in \cref{eq:HS_parameters} and the coupling uses $\phi = \pi/6$.} 
    \label{fig:energy_splittings}
\end{figure}

To further test the bounds on the energy splitting, we couple the end of the chain ($R=\{N\}$) to a spinless quantum dot, acting as the environment, $B$, see \cref{fig:energy_splittings}. Depending on the degree of MBS separation, the ground-state degeneracy splits when tuning the dot level to resonance~\cite{pradaMeasuringMajoranaNonlocality2017,clarkeExperimentallyAccessibleTopological2017,dengMajoranaBoundState2016,soutoProbingMajoranaLocalization2023}.
The idea is analogous to the steps leading to our bounds in \cref{sec:energy_splitting_bounds}: the ground-state degeneracy splits if the dot couples to more than one MBS. 
This protocol was recently tested experimentally in a quantum-dot-based Kitaev chain \cite{bordinProbingMajoranaLocalization2025}.

The Hamiltonian of the coupled chain-dot system is
\begin{equation}
    H = H_K + (t_c f_N f_B^\dagger + \Delta_c f_N f_B + \text{h.c.}) + U_c n_N n_B + \varepsilon_d n_B,
\end{equation}
where $f_B$ is the annihilation operator on the dot, which costs an energy $\varepsilon_d$ to occupy. The coupling consists of tunneling ($t_c$), pairing ($\Delta_c$), and Coulomb interaction ($U_c$) between the dot and the end of the chain. Here, the $t_c$ and $\Delta_c$ terms correspond to the odd coupling $H_{RB}^o$ and the $U_c$ term provides an even coupling. However, parts of the $U_c$ term act only in $B$ (renormalizing $\varepsilon_d$), which we move to $H_B$ to minimize the norm of the coupling. See \cref{app:coupling_hamiltonian_ambiguity_convention} for details on handling ambiguities of $H_{RB}$.

To parametrize $t_c$, $\Delta_c$, and $U_c$, we introduce an overall coupling amplitude $\lambda$ and an angle $\phi$ that tunes between tunneling- and pairing-dominated coupling. In terms of these parameters, we set
\begin{subequations}
\begin{align}
    t_c &= \lambda \cos{\phi}, \\
    \Delta_c &= \lambda \sin{\phi}, \\
    U_c &= \lambda,
\end{align}
\end{subequations}
where we chose odd and even couplings to be of similar magnitude, as observed experimentally~\cite{bordinProbingMajoranaLocalization2025}.
In this case, we can express $\norm{H_{RB}^o}_p$ and $\norm{H_{RB}^e}_p$ in terms of $\lambda$ and $\phi$. In particular, for $q=p=2$, they only depend on $\lambda$. We find $\norm{H_{RB}^e}_2 = \lambda/\sqrt{2}$, and $\norm{H_{RB}^o}_2 = \sqrt{2}\lambda$.

If the isolated chain is degenerate ($\varepsilon=0$), we can simplify the bound in \cref{eq:non_perturbative_bound} to
\begin{equation}\label{eq:non_perturbative_bound_ex}
    |\delta E| \leq \frac{\LD{N}_2 + 2\LF{N}_2 \norm{(|\stateE\rangle\langle\stateO|)_B}_2}{\sqrt{2}|\langle \mathcal{O} | \gamma | \mathcal{E} \rangle|}\lambda ,
\end{equation}
where we inserted the above expressions for $\norm{H_{RB}^e}_2$ and $\norm{H_{RB}^o}_2$. This can be further simplified to the weaker bound
\begin{equation}\label{eq:perturbative_bound_ex}
    |\delta E| \leq \left(\LD{N}_2 + 2\LF{N}_2 \right)\lambda,
\end{equation}
where we used $\norm{(|\stateE\rangle\langle\stateO|)_B}_2 \leq 1$ and $|\langle \mathcal{O} | \gamma | \mathcal{E} \rangle| \geq 1/\sqrt{2}$. We select $|\mathcal{E}\rangle$ and $|\mathcal{O}\rangle$ as the ground states of the coupled system.

In \cref{fig:energy_splittings}, we compare the bounds in \cref{eq:perturbative_bound_ex,eq:non_perturbative_bound_ex} with the actual ground-state energy splitting of the coupled system while varying $\varepsilon_d$. At weak coupling ($\lambda = t/100$), the splitting closely follows that of the effective model, for which the bounds are strictly satisfied. At strong coupling ($\lambda = t/2 \sim$ excitation gap), higher-order contributions become relevant, yet the bounds remain satisfied in this example.

The weaker bound in \cref{eq:perturbative_bound_ex} depends only on the coupling strength and the ground-state MBSs in the chain, which do not depend on $\varepsilon_d$. This bound constrains the energy splitting to be less than $\sim 6\%$ of $\lambda$, due to the quality measures $\norm{\gamma_N}_2 \approx \norm{(i\gamma\tilde{\gamma})_N}_2 \approx 0.02$. 

The stronger bound \cref{eq:non_perturbative_bound_ex} includes the factors $\norm{(|\stateE\rangle\langle\stateO|)_B}_2$ and $|\langle \mathcal{O} | \gamma | \mathcal{E} \rangle|$, which depend on $\varepsilon_d$ since they involve eigenstates of the full system. We found that $|\langle \mathcal{O} | \gamma | \mathcal{E} \rangle| \approx 1$ in this example, while
$\norm{(|\stateE\rangle\langle\stateO|)_B}_2$ accounts for most of the variation of the bound. This factor indicates whether an odd coupling can cause a splitting. It peaks at the resonance where there can be a strong mixing between the states of the system and the environment. 

When $\varepsilon_d$ becomes negative, the bound saturates, but it becomes increasingly less tight for positive $\varepsilon_d$. This is because there is another mechanism protecting the energy splitting, which is not due to localized MBSs. When $\varepsilon_d \gg 0$, the dot $B$ is empty and the Coulomb interaction has no effect.

\section{Example II: Minimal quantum-dot-based Kitaev chain}\label{sec:example_abs_to_ysr}
As a second example, we consider a minimal quantum-dot–based Kitaev chain realized in recent experiments~\cite{dvirRealizationMinimalKitaev2023,tenhaafTwositeKitaevChain2024,zatelliRobustPoorMans2024,bordinEnhancedMajoranaStability2025,tenhaafObservationEdgeBulk2025,bordinProbingMajoranaLocalization2025}. We show that the experimental tuning procedure suppresses only a restricted class of even perturbations, while leaving odd perturbations unconstrained. 
We further connect the odd quality measures to transport signatures and provide bounds on the nonlocal conductance in \cref{sec:transport}.

\subsection{Model}
The setup consists of two quantum dots ($L, R$) coupled via a proximitized hybrid section ($H$) that hosts extended Andreev bound states, see \cref{fig:abs_ysr}(a). Due to strong intradot Coulomb interactions, a many-body description is required. Following Refs.~\cite{tsintzisCreatingDetectingPoor2022,liuEnhancingExcitationGap2024}, which provide a theoretical description of the experimentally realized quantum-dot–based Kitaev chain devices in Refs.~\cite{dvirRealizationMinimalKitaev2023,tenhaafTwositeKitaevChain2024,zatelliRobustPoorMans2024}, we model the system by the Hamiltonian
\begin{equation}\label{eq:exampleII_ham}
\begin{aligned}
    H &= H_D + H_H + H_T, \\
    H_D &= \sum_{\sigma, a=L,R} (\varepsilon_{a} + s_\sigma E_{Z}) n_{a\sigma}
    + U_{D} n_{a\uparrow} n_{a\downarrow}, \\
    H_H &= \varepsilon_H (n_{H\uparrow} + n_{H\downarrow})
    + \Delta_H (f_{H\uparrow} f_{H\downarrow} + \mathrm{h.c.}),\\
    H_T &= \sum_{\sigma} \Bigl(
    t f_{H\sigma}^\dagger f_{L\sigma}
    + s_\sigma \alpha_\mathrm{so} t f_{H\bar{\sigma}}^\dagger f_{L\sigma}\\
    &+ t f_{R\sigma}^\dagger f_{H\sigma}
    + s_\sigma \alpha_\mathrm{so} t f_{R\bar{\sigma}}^\dagger f_{H\sigma}
    + \mathrm{h.c.}\Bigr).
\end{aligned}
\end{equation}
Here, $H_D$ describes the isolated dots with level energies $\varepsilon_L, \varepsilon_R$, Zeeman splitting $E_Z$ ($s_{\uparrow(\downarrow)} = \pm 1$), and intradot Coulomb interaction $U_D$, while $H_H$ captures the low-energy hybrid section with induced pairing $\Delta_H$ and normal-state energy $\varepsilon_H$. Zeeman splitting and Coulomb interactions in the hybrid region are neglected due to superconducting renormalization effects~\cite{mikkelsenHybridizationSuperconductorsemiconductorInterfaces2018,antipovEffectsGateinducedElectric2018}. The tunneling Hamiltonian $H_T$ includes both spin-conserving and spin-flipping processes with amplitudes $t$ and $\alpha_\mathrm{so}t$, respectively, and we assume left-right symmetry. Unless stated otherwise, we fix $E_Z = 1.5\Delta_H$, $U_D = 5\Delta_H$, and $\alpha_\mathrm{so} = 0.3$, consistent with experimental parameters in similar devices~\cite{dvirRealizationMinimalKitaev2023,wangTripletCorrelationsCooper2023,bordinTunableCrossedAndreev2023}.

In the limit $E_Z, \Delta_H \gg t$, and when a single spin-polarized level on each dot is tuned close to zero energy, the low-energy sector of $H$ reduces to an effective two-site Kitaev chain~\cite{liuTunableSuperconductingCoupling2022}, hosting spatially separated MBSs at fine-tuned points in parameter space~\cite{leijnseParityQubitsPoor2012}. 
However, even outside the strict Kitaev limit, parameter regimes with partial protection against local perturbations can be identified~\cite{tsintzisCreatingDetectingPoor2022,liuEnhancingExcitationGap2024}.

\subsection{Experimental tuning and quality measures}
In an experiment, the system would be tuned by varying $\varepsilon_L$, $\varepsilon_R$, and $\varepsilon_H$ while searching for crossings in a charge-stability diagram. These correspond to points where the energy splitting between the odd and even parity ground states, $\delta E$, vanishes and is stationary with respect to dot-level energies,
\begin{equation}\label{eq:crossing_condition}
\partial_{\varepsilon_a} \delta E
= \langle e | n_a | e \rangle - \langle o | n_a | o \rangle
\equiv \delta n_a = 0,
\end{equation}
where $n_a = n_{a\uparrow} + n_{a\downarrow}$ and $a=L,R$. This condition requires vanishing total charge difference on each dot, while allowing for compensating contributions from opposite spin sectors.

By contrast, the spin-resolved even quality measure
\begin{equation}\label{eq:exampleII_Qe}
Q_e
= \sqrt{\sum_{\sigma,a=L,R}
\left\lVert (i\gamma \tilde\gamma)_{a\sigma} \right\rVert_2^2}
\end{equation}
vanishes only if the charge differences vanish separately for each spin component, $\delta n_{a\sigma}=0$. While both conditions protect against dot-level fluctuations, $Q_e=0$ additionally guarantees robustness against Zeeman-like perturbations, which are not constrained by \cref{eq:crossing_condition}.

The odd quality measure
\begin{equation}\label{eq:exampleII_Qo}
Q_o^a = \sqrt{\sum_\sigma \lVert \gamma_{a\sigma} \rVert_2^2},
\end{equation}
quantifies protection against odd local perturbations acting on dot $a=L,R$ (in our case, $Q_o^L = Q_o^R = Q_o$). This aspect of protection is not addressed by experimental tuning and is directly relevant for coupling to external quantum dots, transport signatures, and coupling-based braiding protocols discussed below.

Finally, an even stronger notion of protection is obtained by defining the subregions as entire dots rather than individual spin sectors. The resulting site-resolved quality measures probe a broader class of local perturbations, including those that couple different spin sectors, such as fluctuations in local pairing amplitudes and density-assisted tunneling.

\subsection{Quality measures from weak to strong dot-hybrid coupling}

\begin{figure}[t]
    \centering
    \includegraphics[width=\linewidth]{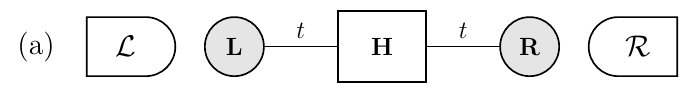}
    \includegraphics[width=\linewidth]{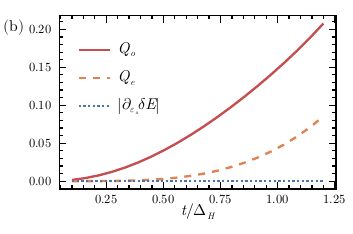}
    \caption{(a) Illustration of a minimal quantum-dot-based Kitaev chain, where two dots $L$ and $R$ are coupled via a hybrid section $H$ and connected to leads $\mathcal{L}$ and $\mathcal{R}$. (b) Even and odd quality measures, $Q_e$ and $Q_o$, and the sensitivity of the energy splitting to dot-level fluctuations, quantified by $|\partial_{\varepsilon_a}\delta E|$ (symmetric in $a=L,R$), shown as a function of the dot-hybrid coupling strength $t$. While the tuning enforces $\partial_{\varepsilon_a}\delta E\approx 0$, both quality measures remain finite, with $Q_o > Q_e$ throughout.}
    \label{fig:abs_ysr}
\end{figure}

In \cref{fig:abs_ysr}(b), we numerically optimize $\varepsilon_L=\varepsilon_R$ and $\varepsilon_H$ to satisfy the experimental tuning condition in \cref{eq:crossing_condition} while maintaining degenerate ground states. This procedure is repeated for a range of coupling strengths $t$. For all $t$, the optimization converges to solutions with
$|\partial_{\varepsilon_a}\delta E|$ and $|\delta E|/\Delta_H$ both less than $10^{-10}$.

At the optimized parameter points, we evaluate the even and odd quality measures $Q_e$ and $Q_o$. Since the tuning enforces a small sensitivity of the energy splitting to dot-level variations, it is more closely related to $Q_e$ than to $Q_o$, which is not directly constrained by the experimental tuning procedure. Accordingly, we find $Q_o>Q_e$ over the full range of coupling strengths shown.
We note that tuning $\varepsilon_{L}, \varepsilon_R$ and $\varepsilon_H$ is not sufficient to drive either $Q_e$ or $Q_o$ to zero while enforcing degenerate ground states.

Increasing the coupling strength is known to enhance the excitation gap~\cite{liuEnhancingExcitationGap2024,zatelliRobustPoorMans2024}. However, the relatively large values of $Q_e$ at strong dot-hybrid coupling indicate that, despite the enforced insensitivity to dot-level variations, the ground-state degeneracy remains sensitive to other local even perturbations. For example, perturbations in the Zeeman energy lead to a linear splitting of the degeneracy, with a magnitude suppressed by $Q_e\approx 0.05$, corresponding to splittings at the few-percent level of the underlying perturbation scale. 

The protection against odd perturbations, as quantified by $Q_o$, is further reduced at larger coupling strengths.
This suggests an alternative route in which experimental tuning is directed toward minimizing 
$Q_o$, for example, by coupling the system to a quantum dot (see \cref{sec:example_coupling}), or via transport measurements, as discussed in the following subsection.

\subsection{Transport signatures of odd protection}\label{sec:transport}
The nonlocal conductance $G_{\alpha\beta} = \partial I_\alpha /\partial \mu_\beta$ measures how the current in  lead $\alpha$ changes with variations in the chemical potential of another lead $\beta$. In noninteracting systems, it is suppressed for spatially separated MBSs and becomes finite when they overlap~\cite{danonNonlocalConductanceSpectroscopy2020,rosdahlAndreevRectifierNonlocal2018}. We show that this connection persists in interacting systems, with the nonlocal conductance being bounded by the odd quality measures.

Using a rate-equation description of transport between the lowest even and odd ground states $|e\rangle$ and $|o\rangle$, we show that the nonlocal differential conductance between two leads $\alpha \neq \beta$ has the form
\begin{equation}
    G_{\alpha\beta} = g \eta_\alpha \eta_\beta,
\end{equation}
see \cref{app:transport} for details. Here, $\eta_\alpha = |\langle o | f_\alpha^\dagger | e \rangle|^2 - |\langle o | f_\alpha | e \rangle|^2$ quantifies the asymmetry between tunneling in and out of the system through lead $\alpha$, and $f_\alpha^\dagger$ describes tunneling from lead $\alpha$ into a region $R_\alpha$ of the system. The prefactor $g$ depends on the temperature and chemical potential of the leads, as well as on the total tunneling rates, but is independent of $\eta_\alpha$ and $\eta_\beta$. Finally, $\eta_\alpha$ can be bounded in terms of $\lVert \gamma_{R_\alpha} \rVert_q$, implying that nonlocal conductance is suppressed when the ground-state MBSs are well separated.

For the experimentally relevant case of two spinful normal leads $\mathcal{L}$ and $\mathcal{R}$ locally coupled to dots $L$ and $R$ (see \cref{fig:abs_ysr}(a)), the bound on the nonlocal conductance can be expressed directly in terms of the odd quality measures,
\begin{equation}
    |G_{LR}| \leq g' Q_o^L Q_o^R,
\end{equation}
where $g'$ depends on temperature and total tunneling rates but not on the asymmetries $\eta$.
Thus, small values of $Q_o$ imply a strong suppression of nonlocal conductance, while finite nonlocal conductance signals imperfect odd-parity protection. Transport measurements, therefore, probe an aspect of protection that is not directly constrained by the experimental tuning procedure. As shown in \cref{fig:abs_ysr}(b), $Q_o$ increases with dot–hybrid coupling, which is precisely the regime where experiments observe a nonzero nonlocal conductance~\cite{zatelliRobustPoorMans2024}.

\section{Braiding ground-state MBS\MakeLowercase{s}}\label{sec:braiding}
\begin{figure}[th]
    \centering
    \includegraphics[width=.8\linewidth]{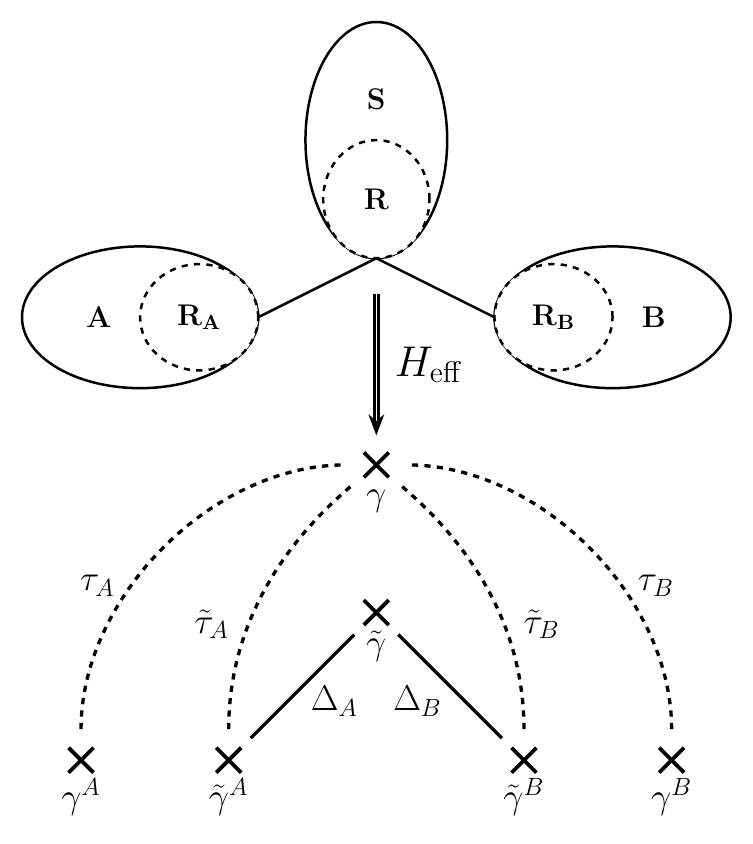}
    \caption{When coupling multiple systems, each with a low-energy fermionic mode, we can formulate an effective theory involving only the ground-state MBSs. With three systems, non-abelian braiding can be implemented if the desired couplings (solid lines) can be tuned and the undesired couplings (dashed lines) stay small. We put bounds on the undesired couplings and specify how to maximize the desired ones.}
    \label{fig:braiding}
\end{figure}

By coupling several systems with MBSs, see \cref{fig:braiding}, one can perform coupling-based braiding protocols \cite{aliceaNonAbelianStatisticsTopological2011, sauControllingNonAbelianStatistics2011, beenakkerSearchNonAbelianMajorana2020} to realize topological quantum computation \cite{nayakNonAbelianAnyonsTopological2008}. In such protocols, certain couplings between the MBSs must remain small to suppress errors~\cite{tsintzisMajoranaQubitsNonAbelian2024, nitschAdiabaticNonabelianBraiding2025, milesBraidingMajoranasLinear2025}, while others must be large to enable the exchanges. In this section, we use our methods to generalize these statements to an interacting setting. We couple three interacting systems and derive an effective Hamiltonian in terms of ground-state MBSs. We then show how the locality of the MBSs bounds unwanted couplings in the braiding protocol. 

A central system $S$ with ground-state MBSs $\gamma$ and $\tilde\gamma$ couples to two external and mutually decoupled systems, $A$ and $B$, with ground-state MBSs $\gamma^A, \tilde\gamma^A$ and $\gamma^B, \tilde\gamma^B$. We assume that the couplings act only between a subregion $R \subset S$ and subregions $R_A \subset A$ and $R_B \subset B$, see \cref{fig:braiding}. The coupling Hamiltonian for this setup is
\begin{equation}
    H_c = H_{RR_A} + H_{RR_B}.
\end{equation}
We consider only the odd part of $H_c$ here, and refer to \cref{app:braiding} for the general case. The effective Hamiltonian is
\begin{equation}\label{eq:braiding_eff}
    \begin{aligned}
       H^o_{c,\text{eff}} = \frac{1}{4}\sum_{X=A,B} 
       \Delta_X i\tilde\gamma \tilde\gamma^X +  \tau_X i\gamma \gamma^X +  \tilde\tau_X i\gamma \tilde\gamma^X,
    \end{aligned}
\end{equation}
where we have picked the gauges of $A$ and $B$ to minimize the number of couplings.
If the coupling varies in time, the effective couplings become time-dependent. A successful braiding protocol can then be performed if the desirable couplings $\Delta_X$ are tunable while the undesirable couplings $\tau_X$ and $\tilde\tau_X$ are small.
The latter ones are bounded by
\begin{subequations}\label{eq:braiding_bounds}
    \begin{align}
        |\tau_X| &\leq \norm{H_{RR_X}^o}_p \norm{\gamma_R}_q \lVert{(\gamma^X)_{R_X}}\rVert_q, \\
        |\tilde\tau_X| &\leq \norm{H_{RR_X}^o}_p \norm{\gamma_R}_q \lVert{(\tilde{\gamma}^X)_{R_X}}\rVert_q.
    \end{align}
\end{subequations}

For the desired couplings to be large, there needs to be MBSs in the subregions that are coupled, i.e., $\norm{\tilde{\gamma}_R}$ should be large, and similarly for the other systems.
Furthermore, $H_{RR_A}$ and $H_{RR_B}$ need to be chosen such that the $\Delta$-terms do not vanish. The $q=2$ bound for $\Delta_X$ is saturated if
\begin{equation}\label{eq:braiding_optimal_coupling}
    H_{RR_X} \propto \tilde\gamma_R (\tilde\gamma^X)_{R_X},
\end{equation}
which is therefore optimal to maximize the desired effective couplings. In an experiment, it may not be feasible to tune the coupling to exactly match the optimal one. However, an imperfect coupling does not ruin the protocol. The desired couplings would be smaller, which increases adiabatic errors, but this can be compensated for by increasing the operation time.

These results provide simple criteria for assessing whether a system is suitable for use in a coupling-based braiding setup: if $\norm{\gamma_R}_q$ and $\norm{(i\gamma\tilde\gamma)_R}_q$ are small, and $\norm{\tilde{\gamma}_R}_q$ is large, then three copies of the system can be used to implement non-abelian braiding.

\section{Conclusions}
For non-interacting systems, the spatial separation of MBSs is an established tool to understand the robustness of the energy spectrum and non-abelian braiding. This concept relies on single-particle wave functions, which are not available in the interacting case.
In this work, we developed a formalism valid in both interacting and non-interacting systems, based on the partial trace of ground-state MBSs, allowing us to use the concept of MBS localization in any system with two low-energy ground states of different fermionic parity. We proved rigorous bounds on the protection of the energy spectrum, and discussed the implications for braiding protocols.
To illustrate our results, we studied the interacting Kitaev chain and found that some of the intuition gained from single-particle MBSs carries over to ground-state MBSs. We further applied our framework to a quantum-dot–based Kitaev chain, showing how experimentally motivated tuning protocols relate to our quality measures and how odd-parity protection can be probed via nonlocal transport signatures.

The partial trace is usually used to study individual states via the reduced density matrix. We point out that when considering multiple states, such as in a topological theory with several (nearly) degenerate ground states, the reduced transition operators are also meaningful and can reveal topological edge states. It would be interesting to understand their structure in more detail. In particular, how the properties of the two ground states (such as their entanglement) influence the properties of the ground-state MBSs. 

Our bounds involve quality measures that generalize those previously used for MBSs, namely the Majorana polarization and the local distinguishability. We clarified the conditions under which previously used formulas remain valid, and when they need to be generalized.
The construction of ground-state MBSs and the quality measures rely only on the ground states, which means our methods can be used even in large interacting systems using, e.g., density matrix renormalization group (DMRG) methods. Thus, our results open up a new way to study strongly interacting topological systems.

We have studied systems with two ground states of opposite parity. It would be interesting to generalize our framework to systems 
containing multiple pairs of MBSs or different types of anyons. 
However, higher degeneracies provide significantly greater freedom in operator parametrization and even in the choice of anyon type \cite{bozkurtInteractioninducedStrongZero2025}, making the analysis more involved.
Another natural generalization would be to consider number-conserving settings, where similar operator constructions of MBSs have been suggested~\cite{ortizManyBodyCharacterizationParticleConserving2014, ortizWhatParticleconservingTopological2016}.

\section{Data availability}
The code and data for this article are available at \cite{samuelsonQuantifyingRobustnessLocality2025}.

\begin{acknowledgments}
We thank Rubén Seoane Souto for helpful discussions. We acknowledge funding from the European Research Council (ERC) under the European Unions Horizon 2020 research and innovation programme under Grant Agreement No. 856526, the European Union’s Horizon Europe research and innovation programme under the Marie Skłodowska-Curie grant agreement No. 101126636, European Innovation Council Pathfinder grant No. 101115315 (QuKiT), Microsoft Research, NanoLund, and the Dutch Research Council (NWO) grant OCENW.GROOT.2019.004.
\end{acknowledgments}

\bibliography{bibliography} 

\appendix

\section{Proof of coupling bounds}
\label{app:general_effective_operator_bounds}
Consider a bipartite system with regions $X,\comp{X}$. We use $\embed{\comp{X}}(\cdot)$ to denote the fermionic embedding of an operator in $X$ into the full space $X\cup \comp{X}$, see Ref.\,\cite{szalayFermionicSystemsQuantum2021}. We denote the fermionic partial trace of an operator $A$ acting in $X\cup\comp{X}$ by $A_X = \Tr_{\comp{X}}[A]$.

The main facts we need to prove the bounds on couplings are
\begin{enumerate}
    \item The Schatten norm satisfies 
\begin{equation}
\norm{B}_p = \sup_{\norm{C}_q=1}{|\Tr[BC]|}.
\end{equation}
\item The fermionic partial trace is the adjoint of the fermionic embedding \cite{szalayFermionicSystemsQuantum2021}, i.e., $\Tr[\embed{\comp{X}}(A)C] = \Tr[A\Tr_{\comp{X}}[C]] = \Tr[A C_X] $.
\item $|\Tr{AB}| \leq \norm{A}_p \norm{B}_q$ for any two matrices $A,B$.
\item If $A_X$ has a specific parity, then
$\embed{X}(A) \embed{\comp{X}}(B)$ is unitarily equivalent to $ A_X \otimes B_{\comp{X}}$ and therefore their Schatten norms are the same. The case when both $A$ and $B$ are even is proved in \cite{szalayFermionicSystemsQuantum2021} by constructing a unitary $U$. This can be extended to the more general case. Assume that $A_X$ is even; otherwise multiply it by an odd unitary. In this case, the unitary that does the job is $ U' = (-1)^{N_X N_{\comp{X}}} U$, where $N_X$ is the number operator in region $X$.
\end{enumerate}

Let $H$ be an operator in the full system, $A$ an operator in $X$ with definite parity, and $B$ any operator in $\comp{X}$. We can bound the trace of their product by
\begin{equation}\label{eq:proof_gapped_bound}
\begin{split}
    |\Tr{\left[\embed{\comp{X}}(A) \embed{X}(B) H\right]}| & \stackrel{3}{\leq} \norm{H}_p \norm{\embed{\comp{X}}(A) \embed{X}(B)}_q \\
    &\stackrel{4}{=} \norm{H}_p \norm{A \otimes B}_q \\
    &=\norm{H}_p \norm{A}_q \norm{B}_q.
\end{split}
\end{equation}
This bound applies to the couplings of gapped systems in \cref{sec:braiding} and \cref{app:braiding}. It is saturated when 
\begin{equation}
    H = \embed{X}((B^\dagger B)^{q/2 -1}B^\dagger)\embed{\comp{X}}((A^\dagger A)^{q/2 - 1} A^\dagger).
\end{equation}
If $q/2<1$ and the matrices are not invertible, one can use the Moore-Penrose pseudoinverse.

Next, we derive the bound used for the effective couplings in \cref{sec:coupling_bounds}. Let $H$ be an operator in the full system, $A$ an operator in $X$ with definite parity and
\begin{equation}\label{eq:B_definition}
    B_{\comp{X}} \equiv \Tr_X\left[\embed{\comp{X}}(A) H\right].
\end{equation}
The effective operators in \cref{eq:effective_ops} have this form. We will prove that 
\begin{equation}
    \norm{B_{\comp{X}}}_p \leq \norm{H}_p \norm{A}_q,
\end{equation}
where $1/p+1/q = 1$. The proof is
\begin{equation} 
\begin{split}
    \norm{B_{\comp{X}}}_p &\stackrel{1}{=} \sup_{\norm{C}_q=1}{|\Tr[BC]|} \\
    &\stackrel{\ref{eq:B_definition}}{=} \sup_{\norm{C}_q=1}{|\Tr[\Tr_X\left[\embed{\comp{X}}(A) H\right]C]|} \\
    &\stackrel{2}{=} \sup_{\norm{C}_q=1}{|\Tr[\embed{\comp{X}}(A)H\embed{X}(C)]|} \\
    &\stackrel{3}{\leq}  \norm{H}_p \sup_{\norm{C}_q=1}{ \norm{\embed{\comp{X}}(A) \embed{X}(C)}_q} \\
    & \stackrel{\ref{eq:proof_gapped_bound}}{\leq} \norm{H}_p \sup_{\norm{C}_q=1}{ \norm{A}_q \norm{C}_q} \\
    &= \norm{H}_p \norm{A}_q.
\end{split}
\end{equation} 
The bound is saturated when
\begin{equation}
    H = \embed{X}(h_{\comp{X}}) \embed{\comp{X}}((A^\dagger A)^{q/2 - 1} A^\dagger) ,
\end{equation}
for any operator $h_{\comp{X}}$ in $\comp{X}$.

\section{Proof of energy splitting bounds}\label{app:energy_splitting_bounds}
This Appendix contains the details on the bound for the energy splitting. The idea is to consider the quantity
\begin{equation}
    \mathcal{C} \equiv [\embed{B}(\gamma), H_\text{eff}] - i\varepsilon\embed{B}(\tilde{\gamma}),
\end{equation}
which, because of the commutator, can be related to energy differences, and can be bounded because $H_\text{eff}$ can be expressed in terms of the effective couplings.

Let $|\stateE\rangle, |\stateO\rangle$ be two eigenstates of $H_\text{eff}$ of even and odd parity. The matrix element
\begin{multline}
    |\langle\stateO|\mathcal{C} | \stateE\rangle| = |\langle\stateO | [\embed{B}(\gamma), H] - i \varepsilon \embed{B}(\tilde{\gamma}) | \stateE\rangle| \\
    = |\delta E \langle \stateO | \embed{B}(\gamma) | \stateE\rangle  - i \varepsilon \langle \stateO | \embed{B}(\tilde{\gamma}) | \stateE\rangle |
\end{multline}
provides us with the energy difference $\delta E$ between the eigenstates. Next, $\mathcal{C}$ can also be expressed in terms of the  effective couplings as
\begin{align}
\mathcal{C} = \embed{S}(G)\embed{B}(i\tilde{\gamma}) + \embed{S}(F).
\end{align}
First, we use the triangle inequality
\begin{multline}
    |\langle \stateO|\mathcal{C}|\stateE\rangle| 
    \leq |\langle\stateO| \embed{B}(\tilde{\gamma}) \embed{S}(G) |\stateE\rangle| + |\langle \stateO|  \embed{S}(F) |\stateE\rangle|,
\end{multline}
and then we will bound the two terms separately. In addition to the properties used in the previous Appendix, we will use that
\begin{enumerate}
    \setcounter{enumi}{4}
    \item The Schatten-1 norm gets smaller under the partial trace \cite{rasteginRelationsCertainSymmetric2012}.
    \item For any two matrices $A$ and $B$ and $1/p + 1/q = 1/r$, we have that $\norm{AB}_r \leq \norm{A}_p \norm{B}_q$.
    \item $\norm{A}_p \leq \norm{A}_q$ for $p>q$.
\end{enumerate}
The $G$-term can be bounded as
\begin{equation}
\begin{split}
    |\langle\stateO | \embed{B}(\tilde{\gamma}) \embed{S}(G) | \stateE\rangle| &= |\Tr[\embed{S}(G) \embed{B}(\tilde{\gamma}) |\stateE\rangle\langle\stateO|]|\\
    & \stackrel{2}{=} |\Tr[G (\embed{B}(\tilde{\gamma}) |\stateE\rangle\langle\stateO|)_B]| \\   
    &\stackrel{3}{\leq} \norm{G}_\infty \norm{(\embed{B}(\tilde{\gamma}) |\stateE\rangle\langle\stateO|)_B}_1 \\ 
    &\stackrel{5}{\leq} \norm{G}_\infty \norm{\embed{B}(\tilde{\gamma}) |\stateE\rangle\langle\stateO|}_1 \\
    &\stackrel{6}{\leq} \norm{G}_\infty \norm{\embed{B}(\tilde{\gamma})}_\infty \norm{|\stateE\rangle\langle\stateO|}_1  \\
    & = \norm{G}_\infty \\
    & \stackrel{7}{\leq} \norm{G}_p.
\end{split}
\end{equation}
The $F$-term is bounded by
\begin{equation}
\begin{split}
    |\langle \stateO|\embed{S}(F)|\stateE\rangle| &\leq  \norm{F}_p\norm{(|\stateE\rangle\langle\stateO|)_B}_q,
\end{split}
\end{equation}
by using properties 2 and 3.

Combining these statements, we end up with the general bound in the main paper
\begin{equation}
\begin{split}
    &|\delta E\langle \stateO | \embed{B}(\gamma) | \stateE\rangle - i\varepsilon \langle \stateO | \embed{B}(\tilde{\gamma}) | \stateE\rangle | 
     \\
    &\leq \norm{G}_p  + \norm{F}_p \norm{(|\stateE\rangle\langle\stateO|)_B}_q 
    \\
    &\leq \norm{(i\gamma\tilde{\gamma})_R}_q \norm{H_{RB}^e}_p + \norm{\gamma_R}_q \norm{H_{RB}^o}_p \norm{(|\stateE\rangle\langle\stateO|)_B}_q.
\end{split}
\end{equation}

\subsection{Simplifying the bound} 
Assume that when the systems are uncoupled, the environment has a cluster of $d_B$ eigenstates which are separated from other eigenstates by an energy gap $E_B$. We consider the case $\norm{H_{RB}}\ll E_B$ so that we can neglect any mixing with other states. We assume that the cluster consists of $d_B^e$ states with even parity and $d_B^o$ states with odd parity, and denote them by $|\psi_j^e\rangle$ and $|\psi_j^o\rangle$.

When the coupling between the system and the bath is turned on, the eigenstates can be expressed as
\begin{align}\label{eq:coupled_eigenstates}
    |\stateE_i\rangle &= \sum_{j=1}^{d_B^e} a_{ij}|e\rangle|\psi_j^e\rangle + \sum_{j=1}^{d_B^o} b_{ij}|o\rangle |\psi_j^o\rangle ,\\
    | \stateO_i\rangle &= \sum_{j=1}^{d_B^e} c_{ij}|o\rangle |\psi_j^e\rangle + \sum_{j=1}^{d_B^o} d_{ij}|e\rangle |\psi_j^o\rangle.
\end{align}

\subsubsection{No parity mixing} 
In general, the states in \cref{eq:coupled_eigenstates} mix states with different parities in $S$ and $B$. If that does not happen, we can simplify things considerably. This simplification happens, for example, if $\varepsilon \gg \norm{H_{RB}}$ or if either $d_B^e=0$ or $d_B^o=0$. Then one can treat the different parity sectors in $B$ independently. Here, we consider the even parity states of the environment. The eigenstates take the form
\begin{align}\label{eq:coupled_eigenstates_no_parity_mixing}
    |\stateE_i\rangle &= \sum_{j=1}^{d_B^e} a_{ij}|e\rangle |\psi_j^e\rangle \\
    |\stateO_i\rangle &= \sum_{j=1}^{d_B^e} c_{ij}|o\rangle |\psi_j^e\rangle.
\end{align}
It follows that
\begin{align}
    &\norm{(|\stateE_i\rangle\langle\stateO_j|)_B}_q = 0 \\
    &\langle\stateO_i | \gamma  | \stateE_j\rangle = i \langle\stateO_i | \tilde{\gamma} | \stateE_j\rangle
\end{align}
for all pairs. The bound simplifies to
\begin{equation}
    \left|\delta E - \varepsilon \right| \left|\langle \stateO | \gamma | \stateE\rangle\right| 
    \leq \norm{(i\gamma\tilde{\gamma})_R}_q \norm{H_{RB}^e}_p.
\end{equation}
Next, we want a lower bound on $\left|\langle\stateO | \gamma | \stateE\rangle\right|$. Since $\gamma |\stateE_i\rangle$ is a normalized state which can be expanded in the basis $|\stateO_j\rangle$, it must overlap with at least one of those states, and that overlap must be larger than $\sqrt{1/d_B^e}$. We arrive at the following statement: for all eigenstates in this set, there exists another eigenstate such that their energy difference is bounded by
\begin{equation}
\lvert \delta E - \varepsilon  \rvert \leq  \LD{R}_q \norm{H_{RB}^e}_p \sqrt{d_B^e}.
\end{equation}
An analogous bound holds when considering odd states of the environment.

\subsubsection{Parity mixing}
In this case, the bound is weaker. Use the triangle inequality on the left-hand side to separate $\delta E$ and $\varepsilon$ and then 
\begin{align}
    &\norm{(|\stateE_i\rangle\langle\stateO_j|)_B}_q \leq 1, \\
    &\left|\langle\stateO_i | i\tilde{\gamma} | \stateE_j\rangle \right| \leq 1, \\
    &\left|\langle\stateO_i | \gamma  | \stateE_j\rangle \right| \geq  \sqrt{1/d_B},
\end{align}
to bound the unknown quantities. 
The result is:
For each eigenstate, there exists another eigenstate with different parity, and their energy difference is bounded by
\begin{equation}\label{eq_app:simple_energy_bound_single_region}
    |\delta E| \leq \sqrt{d_B}(\LD{R}_q \norm{H_{RB}^e}_p + \LF{R}_q \norm{H_{RB}^o}_p +|\varepsilon|).
\end{equation}

\subsection{Uncoupled eigenstates}
A simpler bound is obtained by considering the original uncoupled eigenstates. The energy difference between $|e \rangle |\psi\rangle$ and $|o \rangle|\psi\rangle$, where $|\psi\rangle$ is any state in the environment is 
\begin{multline}
    \delta E \equiv \langle \psi|\langle e | H_\text{eff} |e \rangle |\psi\rangle -  \langle \psi|\langle o | H_\text{eff} |o \rangle |\psi\rangle 
     \\  = \langle \psi|\langle o | [\gamma, H_\text{eff}]  | e \rangle|\psi\rangle.
\end{multline}
Together with \cref{eq:effective_commutator_interacting,eq:effective_operators_bound} it is straightforward to show that
\begin{equation}
    |\delta E - \varepsilon| \leq \LD{R}_q \norm{H_{RB}^e}_p,
\end{equation}
which says that the perturbation of the energy difference only changes if the MBSs overlap in the region $R$. This generalizes a bound from Ref.~\cite{svenssonQuantumDotBased2024}.

\subsection{Bounds for states paired up by $\gamma$}
The bound in \cref{eq_app:simple_energy_bound_single_region} scales with the dimension of the environment, which makes it unsuitable in the case where the environment has a dense spectrum. This factor arises because we compare the energy of two exact eigenstates, which may not be perfectly mapped to each other by $\gamma$. Taking instead an arbitrary state $|\psi\rangle$, we can show that the state $\gamma|\psi\rangle$ will have a similar energy distribution as $|\psi\rangle$. 

First, we bound the commutator by 
\begin{equation}
\begin{aligned}
    \norm{[\gamma, H_\mathrm{eff}]}_p &\leq \norm{G\tilde{\gamma}}_p + \norm{F}_p  + \norm{\tilde{\gamma}}_p|\varepsilon| \\ 
    &\leq \LD{R}_q \norm{H_{RB}^e}_p + \LF{R}_q \norm{H_{RB}^o}_p  \\
     & \quad + (2d_B)^{1/p}|\varepsilon|.
\end{aligned}
\end{equation}
This then bounds the commutators of higher powers of $H_\mathrm{eff}$ as
\begin{equation}
\begin{aligned}
    \norm{[\gamma, H_\mathrm{eff}^n]}_p &= \norm{\sum_{k=0}^{n-1} H_\mathrm{eff}^k [\gamma, H_\mathrm{eff}] H_\mathrm{eff}^{n-1-k}}_p \\
    & \leq n\norm{[\gamma, H_\mathrm{eff}] H_\mathrm{eff}^{n-1}}_p \\
    & \leq n\norm{[\gamma, H_\mathrm{eff}]}_p \norm{H_\mathrm{eff}}^{n-1}_\infty.
\end{aligned}
\end{equation}

Finally, the difference in the $n$:th energy moment between the two states is bounded as
\begin{equation}\label{eq_app:dense_spectrum_bound}
\begin{aligned}
    |\langle\psi|H_\mathrm{eff}^n|\psi\rangle - &\langle\psi|\gamma H_\mathrm{eff}^n\gamma|\psi\rangle| 
     = |\langle\psi|[\gamma, H_\mathrm{eff}^n]\gamma|\psi\rangle| \\ 
    &\leq \norm{[\gamma, H_\mathrm{eff}^n]}_\infty \\
    &\leq n\norm{[\gamma, H_\mathrm{eff}]}_\infty\norm{H_\mathrm{eff}}^{n-1}_\infty \\
    &\leq \Big(\LD{R}_1 \norm{H_{RB}^e}_\infty + \LF{R}_1 \norm{H_{RB}^o}_\infty  \\
     & \qquad + |\varepsilon|\Big)n\norm{H_\mathrm{eff}}^{n-1}_\infty.
\end{aligned}
\end{equation}
where we set $q=1, p=\infty$ to get rid of the factor of $d_B$. This form of the bound applies even if the spectrum is dense.

\section{Choice of norms}\label{app:normchoice}
Throughout this article, we use the Schatten norm and express the bounds for general $p,q$, as different choices have different advantages. The two most interesting cases are $q=1, p=\infty$ (trace and operator norm), and $p=q=2$ (Frobenius norm). 

The choice $q=1$ has the advantage that the partial trace is contractive in the trace norm, i.e.,
\begin{equation}
    R \subset R' \Rightarrow \norm{\gamma_R}_1 \leq \norm{\gamma_{R'}}_1,
\end{equation}
which intuitively says that a region contained in another region must contain less of the MBS. This is not necessarily the case for other norms.
Additionally, if the environment is expanded $B\rightarrow B\cup B'$ trivially as $H_{RBB'} = H_{RB}\otimes I_{B'}$, then the norm is $\norm{H_{RB'}}_p = d_{B'}^{1/p}\norm{H_{RB}}_p$. Choosing $p=\infty$ makes the norm insensitive to this trivial extension.

Finally, for $q=1$, we have that
\begin{equation}\label{eq:parity_LD}
    \norm{(i\gamma\tilde\gamma)_R}_1 = \norm{\rho^e_R - \rho^o_R}_1,
\end{equation}
which is proportional to the trace distance between the reduced density matrix of the even (odd) ground states $\rho_R^{e(o)}$. This distance measure has an operational meaning in that it gives maximum probability to distinguish between the states by a measurement \cite{Nielsen_Chuang_2010}.

On the other hand, choosing the Frobenius norm $p=q=2$ has the advantage that it is easier to evaluate analytically, and optimizing with this norm provides a natural gauge for defining MBSs. While the partial trace in this norm is not necessarily contractive, it satisfies 
\begin{equation}
    \norm{\gamma_R}^2_2 + \norm{\gamma_{\comp{R}}}^2_2 \leq 2\norm{\gamma}^2_2 = 4,
\end{equation}
see Ref.\,\cite{ricoNewPartialTrace2025}.

\section{Ambiguities of the coupling Hamiltonian}\label{app:coupling_hamiltonian_ambiguity_convention}
Writing the coupled Hamiltonian as
\begin{equation}
    H = H_S + H_{RB} + H_B,
\end{equation}
introduces some ambiguities, as
\begin{enumerate}
    \item terms acting only in $R$ can be in either $H_S$ or $H_{RB}$, and
    \item terms acting only in $B$ can be in either $H_B$ or $H_{RB}$.
\end{enumerate}
To minimize the norm of $H_{RB}$, one can put the second type of terms in $B$. In other words, $\Tr_R[H_{RB}] = 0$. One can also put the first type of terms in $H_S$ for the same reason, but note that this will influence the eigenstates of $S$ and its ground-state MBS. In \cref{sec:example}, such terms arise from the Coulomb interaction between $R$ and $B$, and we choose to keep them in $H_{RB}$ to keep the MBSs constant and the degeneracy $\varepsilon=0$.

\section{Coupling to several subregions}\label{app:sumofterms}
Here, we consider the case where the coupling Hamiltonian consists of a sum of terms, each acting in a region $R_n$. These regions may overlap. One can directly use the bounds in the main paper with the region $R = \bigcup_n R_n$. However, one can also treat each region independently. We do so here, defining effective operators for each region, bounding them, and generalizing the bound for the energy spectrum. 

Let the coupling Hamiltonian be
\begin{equation}
    H_{SB} = \sum_n H_{R_nB}.
\end{equation}
The effective Hamiltonian can be written
\begin{equation}
    H_\text{eff} = \varepsilon i\gamma \tilde{\gamma}/2 + \sum_n G_n i\gamma \tilde{\gamma}/2 + \gamma F_n/2 + \tilde{\gamma} \tilde{F}_n/2 + H_B,
\end{equation}
The effective operators are
\begin{subequations}
\begin{align}
    G_n &= \Tr_S{\left[i\gamma\tilde{\gamma} H_{R_nB}^e\right]} = \Tr_R{\left[(i\gamma\tilde{\gamma})_{R_n} H_{R_nB}^e\right]}, \\
    F_n &= \Tr_S{\left[\gamma H_{R_nB}^o\right]} = \Tr_R{\left[\gamma_{R_n} H_{R_n B}^o\right]}, \\
    \tilde{F}_n &= \Tr_S{\left[\tilde{\gamma} H_{R_nB}^o\right]} = \Tr_R{\left[\tilde{\gamma}_{R_n} H_{R_nB}^o\right]}
\end{align}
\end{subequations}
and they are bounded as
\begin{subequations}\label{eq:effective_operators_bound_sum_of_terms}
\begin{align}
    \norm{F_n}_p &\leq \norm{H_{R_nB}^o}_p \norm{\gamma_{R_n}}_q, \\
    \norm{\tilde{F}_n}_p &\leq \norm{H_{R_nB}^o}_p \norm{\tilde{\gamma}_{R_n}}_q, \\
    \norm{G_n}_p &\leq \norm{H_{R_nB}^e}_p \norm{(i\gamma\tilde{\gamma})_{R_n}}_q. 
\end{align}
\end{subequations}
The commutator is 
\begin{equation}
    [\gamma, H_\text{eff}] - \varepsilon i\tilde{\gamma} = \sum_n i\tilde{\gamma}G_n  +  F_n.
\end{equation}
The generalization of the energy bound is
\begin{multline} \label{eq:non_perturbative_bound_sum}
    |\delta E \langle\stateO | \gamma | \stateE\rangle - \varepsilon \langle\stateO | i\tilde{\gamma} | \stateE\rangle | \leq \sum_n \norm{(i\gamma\tilde{\gamma})_{R_n}}_q \norm{H_{R_nB}^e}_p \\
    + \norm{\gamma_{R_n}}_q \norm{H_{R_nB}^o}_p \norm{(|\stateE\rangle\langle\stateO|)_B}_q.
\end{multline}
We make the same simplifications as in \cref{app:energy_splitting_bounds}, and also use Cauchy-Schwartz for the sum on the right-hand side to find
\begin{equation} \label{eq:simple_general_energy_bound_app}
    |\delta E| \leq \sqrt{d_B}\left(Q_e h_e + Q_o h_o
     + |\varepsilon|\right),
\end{equation}
where 
\begin{subequations}
\begin{align}
    Q_o &= \sqrt{\sum_{n} \norm{\gamma_{R_n}}_q^2},\\
    Q_e &= \sqrt{\sum_{n}\norm{(i\gamma \tilde{\gamma})_{R_n}}_q^2}, \\
    h_o &= \sqrt{\sum_n \norm{H_{R_{n}B}^o}_p^2}, \\
    h_e &= \sqrt{\sum_n \norm{H_{R_{n}B}^e}_p^2}.
\end{align}
\end{subequations}
$Q$ is an MBS quality factor which only depends on the subregions in which the coupling acts, and $h$ measures the size of the coupling.

The bound in \cref{eq_app:dense_spectrum_bound} generalizes as
\begin{equation}
\begin{aligned}
    |\langle\psi|H_\mathrm{eff}^n|\psi\rangle - &\langle\psi|\gamma H_\mathrm{eff}^n\gamma|\psi\rangle| 
     = |\langle\psi|[\gamma, H_\mathrm{eff}^n]\gamma|\psi\rangle| \\ 
    &\leq \left(Q_e h_e + Q_o h_o
     + |\varepsilon|\right)n\norm{H_\mathrm{eff}}^{n-1}_\infty,
\end{aligned}
\end{equation}
where one should calculate $Q$ and $h$ using $q=1, p=\infty$.


\section{Details on gauge choices and Majorana polarization}\label{app:optim_gauge}
A shift $\theta$ of the relative phase between $|e\rangle$ and $|o\rangle$ results in a rotation of the ground-state MBSs as 
\begin{subequations}
\begin{align}
    \gamma^\theta &= \cos{\theta}\gamma + \sin{\theta}\tilde{\gamma}, \\
    \tilde{\gamma}^\theta &= -\sin{\theta}\gamma + \cos{\theta}\tilde{\gamma}.
\end{align}
\end{subequations}
In the Frobenius norm ($q=2$), we can analytically derive the optimal gauge that minimizes the norm of one of the MBSs in a region. Another useful feature of the Frobenius norm is that the other MBS has a maximum in that gauge (this is not true in general for other norms). Furthermore, for a real Hamiltonian, the gauge with either real (or imaginary) states is optimal. 

It is convenient to work with the ground-state fermion 
\begin{equation}
    c = (\gamma + i\tilde{\gamma})/\sqrt{2}, 
\end{equation}
which transforms as $c\rightarrow e^{i\theta}c$ under the gauge. The minimum can then be determined as
\begin{equation}\label{eq:gauge_optimization_derivation}
\begin{split}
    \min_\theta \sum_n \norm{\gamma^\theta_{R_n}}_2^2 &= \min_\theta \sum_n \norm{e^{i\theta} c_{R_n}+e^{-i\theta}c_{R_n}^\dagger}_2^2/2 \\
    &= \min_\theta \sum_n \Tr[c^\dagger_{R_n} c_{R_n} ] + \\
    &\qquad \Re\left(e^{2i\theta}\sum_n\Tr[c_{R_n}c_{R_n}]\right) \\
    &= \sum_n \Tr[c^\dagger_{R_n} c_{R_n}] - \left| \sum_n \Tr[c_{R_n}c_{R_n}]\right|,
\end{split}
\end{equation}
and the optimal phase is $\pi/2$ minus half the phase of the number $\sum_n\Tr[c_{R_n}c_{R_n}]$. The first term in \cref{eq:gauge_optimization_derivation} can be interpreted as a fermion density, and the second term as a Majorana density. This can also be written in the form
\begin{equation}\label{eq:LF_int_app}
    \min_\theta \sum_n \norm{\gamma^\theta_{R_n}}_2^2 = \left(\sum_n \norm{c_{R_n}}_2^2\right)(1-M),
\end{equation}
where
\begin{equation}\label{eq:MP_definition_app}
    M = \max_\theta \frac{\sum_n\norm{\gamma_{R_n}^\theta}_2^2-\norm{\tilde{\gamma}_{R_n}^\theta}_2^2}{\sum_n\norm{\gamma_{R_n}^\theta}_2^2+\norm{\tilde{\gamma}^\theta_{R_n}}_2^2} = \frac{\left|\sum_n \Tr[c_{R_n} c_{R_n}] \right|}{\sum_n \Tr[c_{R_n}^\dagger c_{R_n}]}.
\end{equation}
In the case of a single region $R$, \cref{eq:LF_int_app} shows that $\min_\theta \lVert\gamma^\theta_R\rVert_2$ vanishes if $R$ contains no MBS at all ($\norm{c_R}_2 = 0$), or if it contains exactly one MBS ($M = 1$).

The Majorana polarization can be calculated using MBSs defined in any gauge using the expression
\begin{equation}\label{eq:gauge_invariant_majorana_polarization}
    M = \frac{\sqrt{\left(\sum_n \norm{\gamma_{R_n}^\theta}_2^2-\norm{\tilde{\gamma}^\theta_{R_n}}_2^2\right)^2 + 4\left\lvert\sum_n \Tr[\gamma_R^\theta \tilde{\gamma}_{R_n}^\theta ]\right\rvert^2}}{\sum_n \norm{\gamma_{R_n}^\theta}_2^2+\norm{\tilde{\gamma}_{R_n}^\theta}_2^2}.
\end{equation}
Comparing this expression with \cref{eq:MP_definition_app}, we see that $\sum_n\Tr[\gamma^\theta_{R_n} \tilde{\gamma}_{R_n}^\theta] = 0$ in the optimal gauge.

\section{Bounds for non-interacting systems}\label{app:non-interacting_bounds}
\subsection{Couplings}\label{app:non_interacting_bounds_couplings}
Now we assume that all terms which act non-trivially on the system $S$ are at most quadratic in the fermions of $S$. The environment is allowed to be interacting. In this case, we can work in the full Hilbert space rather than at the ground-state level. Begin by diagonalizing $H_S$ in terms of standard Majoranas $\Gamma$, $\tilde\Gamma$ and $\{\chi_n,\tilde\chi_n\}$ as
\begin{equation}\label{eq:HS_nonint}
    H_S = i\varepsilon \Gamma\tilde\Gamma/2 + \sum_{n=1}^{N_S-1} i\varepsilon_n \chi_n\tilde\chi_n/2,
\end{equation}
where $N_S$ is the number of fermionic modes in $S$, and $\varepsilon, \varepsilon_n \geq 0$ are energies corresponding to a pair of Majoranas. In writing \cref{eq:HS_nonint}, we have picked out a particular pair of Majoranas $\Gamma$ and $\tilde{\Gamma}$. These Majoranas could, for instance, be the pair with the lowest energy. Then, $\varepsilon$ is the energy splitting of the ground states in the isolated system $S$, analogously to the interacting case. 

The coupling can be decomposed as
\begin{equation}
    H_{RB} = H_{RB}^e +H_{RB}^o,
\end{equation}
where $H_{RB}^e$ is quadratic in $R$ and $H_{RB}^o$ has terms with one fermion in $R$. The full Hamiltonian can be written
\begin{equation}\label{eq:non_interacting_hamiltonian}
\begin{aligned}
    H &= i(\varepsilon/2 + G) \Gamma \tilde\Gamma + \Gamma F + \sum_{n=1}^{N_S -1} (iG_n \Gamma \chi_n + i\tilde G_n \Gamma \tilde\chi_n) \\ 
    &+ \text{terms commuting with $\Gamma$}, 
\end{aligned}
\end{equation}
where $F, G, G_n, \tilde{G_n}$ are operators acting on the environment, given by
\begin{subequations}
\begin{align}
    F & = \Tr_R[\Gamma_R H_{RB}^o]/d_S,  \\ 
    G & = 
    \Tr_{R}[ i[\Gamma_R,\tilde{\Gamma}_{R}] H^e_{RB}]/ (2d_{\comp{R}}d_S),\\
    G_n &= \Tr_{R}[ i[\Gamma_R,\chi_{nR}] H^e_{RB}]/ (2d_{\comp{R}}d_S),
\end{align}
\end{subequations}
and similarly for $\tilde G_n$.
Here we have used that for two anti-commuting single-particle operators, $(\Gamma\chi)_R = [\Gamma_R, \chi_{R}]/(2d_{\comp{R}})$. If the reduced Majorana $\Gamma_R$ is small, the effective operators will be small. More precisely,
\begin{subequations}
\begin{align}
    \norm{F}_p &\leq \norm{H^o_{RB}}_{p} \norm{\Gamma_R}_q/d_S, \\
    \norm{G}_p & \leq \norm{H^e_{RB}}_{p} \norm{(i\Gamma\tilde{\Gamma})_R}_q/(d_S) \\
    & = \norm{H^e_{RB}}_{p} \norm{[\Gamma_R,\tilde{\Gamma}_{R}]}_q  /(2d_{\comp{R}}d_S), \nonumber \\
    \norm{G_n}_p &\leq \norm{H^e_{RB}}_{p} \norm{[\Gamma_R,\chi_{nR}]}_q  /(2d_{\comp{R}}d_S) \nonumber \\
     &\leq \norm{H^e_{RB}}_{p} \norm{\Gamma_R}_q d_R^{1/q}/d_S,
\end{align}
\end{subequations}
and similarly for $\tilde G_n$. 

Optimizing the gauge to minimize $\norm{\Gamma}_q$, we have 
\begin{equation}
    \min_\text{gauges}\norm{\Gamma_R}_q = \norm{C_R}_q\sqrt{1-M},
\end{equation}
where $C = (\Gamma + i\tilde{\Gamma})/\sqrt{2}$ and 
\begin{equation}
    M = \frac{\bigl\lvert\Tr[C_{R} C_{R}]\bigr\rvert}{\Tr[C^\dagger_{R} C_{R}]}.
\end{equation}
We can also write $\norm{[\Gamma_R,\tilde{\Gamma}_{R}]}_q$ in a similar form. It is gauge independent, so we choose the gauge where $\Gamma_R$ and $\tilde{\Gamma}_{R}$ anticommute. Then one finds that
\begin{equation}
    \norm{[\Gamma_R,\tilde{\Gamma}_{R}]}_q = 2\frac{\norm{C_R}^2_q}{ d_R^{1/q}}\sqrt{1-M^2}
\end{equation}

\subsection{Energy bounds}\label{app:non_interacting_bounds_energy}
We will bound the energy difference between two arbitrary eigenstates $| \stateE\rangle, | \stateO\rangle$ of the full Hamiltonian. As in the interacting case, the strategy is to consider the quantity
\begin{equation}\label{eq:non_interacting_commutator}
\begin{aligned}
    \left[\Gamma, H\right] - i\varepsilon\tilde{\Gamma} = \mathcal{C},
\end{aligned}
\end{equation}
where $\mathcal{C} =  2iG\tilde{\Gamma} + 2F + 2i\sum_n G_n\chi_n + \tilde{G}_n \tilde{\chi}_n$. All terms in $\mathcal{C}$ are small if $\norm{\Gamma_R}$ is small. Repeating the derivation for the energy bound in the interacting case, we find
\begin{equation}
\begin{split}
    &|\delta E\langle\stateO | \Gamma | \stateE\rangle - i\varepsilon \langle\stateO | \tilde{\Gamma} | \stateE\rangle |/2 
     \\
    &\leq \norm{G}_p + \sum_{n=1}^{N_S -1} \norm{G_n}_p+\lVert{\tilde{G}_n\rVert}_p + \norm{F}_p \norm{(|\stateE\rangle\langle\stateO|)_B}_q.
\end{split}
\end{equation}
Using the optimal gauge and the coupling bounds, this becomes 
\begin{equation}
\begin{split}
&|\delta E\langle\stateO | \Gamma | \stateE\rangle - i\varepsilon \langle\stateO | \tilde{\Gamma} | \stateE\rangle |/2 
     \\
     &\leq \frac{\norm{C_R}_q}{d_S}\sqrt{1-M}\Big[\norm{H_{RB}^o}_p\norm{(|\stateE\rangle\langle\stateO|)_B}_q  \\  & \qquad+ \norm{H_{RB}^e}_p\Big( 2(N_S-1)d_R^{1/q} + \frac{\norm{C_R}_q}{d_{\comp{R}} d_R^{1/q}}\sqrt{1+M}\Big) \Big].
 \end{split}
\end{equation}
This expression still contains quantities which are hard to calculate, namely $|\langle\stateO | \Gamma | \stateE\rangle |,|\langle\stateO | \tilde{\Gamma} | \stateE\rangle |$ and $\norm{(|\stateE\rangle\langle\stateO|)_B}_q$. Bounding these is more complicated than for the effective theory of the ground-state sector, as we have to deal with the full Hilbert space of the system. On the other hand, these bounds hold for arbitrarily large coupling. The main takeaway from this equation is the prefactor $\norm{C_R}_q \sqrt{1-M}$, which characterizes the MBS quality.

\subsection{Majorana polarization}
So far, we have worked with the single-particle MBSs as operators in Hilbert space. Since they are single-particle operators, it is also useful to represent them in a single-particle operator basis and study their wavefunction.

We need some notation for converting single-particle operators to their wavefunction. For an operator expanded in a basis of single-particle Majorana operators
\begin{align}
    \chi = \sum_{js} a_{js} \Gamma_j^s,
\end{align}
we define $|\Gamma) = \vec{a}$ to be the vector of the coefficients. By $|\Gamma)_R$ we mean the restriction to only those components $\Gamma_j^s$ that live in the region $R$. It is related to the partial trace by $|\Gamma_R) = d_{\comp{R}} |\Gamma)_R$. The inner product of two single-particle operators in the Hilbert space is related to the inner product of the wavefunctions as
\begin{align}
    \Tr[\chi_1^\dagger \chi_2] &= (\chi_1|\chi_2) d_H \\
    \{\chi_1,\chi_2 \} &= 2(\chi_1|\chi_2) I.
\end{align}

The Majorana polarization \cref{eq:gauge_invariant_majorana_polarization} is
\begin{equation}
     M= \frac{\sqrt{ \left(({\Gamma}|\Gamma)_R-(\tilde{\Gamma}|\tilde{\Gamma})_R\right)^2 + 4 (\tilde{\Gamma}|\Gamma)_R^2}}{({\Gamma}|\Gamma)_R + (\tilde{\Gamma}|\tilde{\Gamma})_R}
\end{equation}

Consider two states $|e\rangle, |o\rangle$ such that $\Gamma|e\rangle = |o\rangle$ and $\tilde{\Gamma}|e\rangle = i|o\rangle$. Correlations of these states define the wavefunctions as
\begin{align}
    (\Gamma|\Gamma)_R &= \sum_{s,j\in R} \Re[\langle o|\Gamma_j^s|e\rangle ]^2 \\
    (\tilde{\Gamma}|\tilde{\Gamma})_R &= \sum_{s,j\in R} \Im[\langle o|\Gamma_j^s|e\rangle ]^2 \\
    (\tilde{\Gamma}|{\Gamma})_R &= \sum_{s,j\in R} \Re[\langle o|\Gamma_j^s|e\rangle ]\Im[\langle o|\Gamma_j^s|e\rangle]
\end{align}
and then the Majorana polarization can be expressed as
\begin{align}
    M = \frac{\left| \sum_{s,j\in R} \langle o|\Gamma_j^s|e\rangle ^2 \right|}{ \sum_{s,j\in R} |\langle o|\Gamma_j^s|e\rangle |^2}.
\end{align}

The structure of these equations becomes clearer if we define the vector $\mathbf{z}_R$ with elements 
\begin{equation}
    (\mathbf{z}_R)_{js} = \langle o|\Gamma_j^s|e\rangle,
\end{equation}
where $j\in R$, $s=\pm$. The real and imaginary parts of this complex vector give the coefficients of the two MBSs, and the Majorana polarization is
\begin{align}
    M &= \frac{|\mathbf{z}_R^T \mathbf{z}_R|}{\mathbf{z}_R^\dagger \mathbf{z}_R} 
\end{align}
The two quantities $\mathbf{z}_R^\dagger\mathbf{z}_R$ and $|\mathbf{z}_R^T \mathbf{z}_R|$ are in fact the only two gauge-invariant quantities that can be written down for a region $R$ in a non-interacting theory. Such a quantity must be independent of the basis $\{\Gamma_i^s\}$ and of the relative phase between the odd and even states. A change of basis corresponds to a Bogoliubov rotation of $\mathbf{z}_R$ by a real orthogonal matrix, and the relative phase multiplies $\mathbf{z}_R$ by a phase. The only two invariants of these transformations are $\mathbf{z}_R^\dagger\mathbf{z}_R$ and $|\mathbf{z}_R^T \mathbf{z}_R|$.

\section{Many-body description of Majorana bound states}\label{app:manybody_desc}
The ground-state MBSs $\gamma$ and $\tilde\gamma$ defined in \cref{eq:ground_state_majoranas_definition} are very different from the standard, single-particle MBSs $\Gamma$ and $\tilde\Gamma$ from non-interacting systems. In this Appendix, we contrast the structure and locality of the two types of MBSs, and for non-interacting systems, we demonstrate their relationship. Unlike single-particle MBSs, $\gamma$ and $\tilde\gamma$ are non-local and act non-trivially on the entire system. Nevertheless, when coupling to them in a region $R$, they behave effectively as local operators, captured by $\gamma_R$ and $\tilde\gamma_R$.

In the non-interacting case, we diagonalize $H_S$ in terms of single-particle Majoranas as in \cref{eq:HS_nonint}. Focusing on the low-energy MBS $\Gamma$, we expand it in a position basis as
\begin{equation}\label{eq:non-interacting_pos}
    \Gamma = \sum_{j,s} a_{js} \Gamma_j^s,
\end{equation}
where $a_{js}\in \mathbb{R}$, and $\Gamma_j^s$ form the single-particle Majorana basis defined in \cref{eq:sp_majorana_basis}.

Filling the mode defined by $\Gamma$ and $\tilde\Gamma$ always costs an energy $\varepsilon$, since without interactions it is independent of the filling of the other modes. Therefore, the many-body spectrum is "strong" in the sense that every state in the even sector is paired up with a state in the odd sector separated by $\varepsilon$. $\Gamma$ and $\tilde\Gamma$ map between the odd and even states throughout the spectrum and can be represented in the energy basis as 
\begin{equation}\label{eq:energy_basis_non-interacting}
    \Gamma = |e\rangle\langle o| + \sum_i |e_i\rangle\langle o_i| + \text{h.c.},
\end{equation}
where $i$ indexes the excited states. It is important to choose the relative phases between different sectors correctly to get a single-particle operator \cite{kellsMultiparticleContentMajorana2015}.

Unlike the non-interacting case, where degenerate ground states imply degeneracy between excited states as well, the spectrum in interacting systems is, in general, weak---the ground states could be degenerate while excited states are not. This fact, along with our focus on adiabatic protocols, further motivates our definition of MBSs acting only in the ground-state sector
\begin{equation}\label{eq:energy_basis_interacting}
    \gamma = |e\rangle\langle o| + \text{h.c.}
\end{equation}
While \cref{eq:energy_basis_non-interacting} and \cref{eq:energy_basis_interacting} have a similar form in the energy basis, the positional representation of $\gamma$ is more complicated than that of $\Gamma$, as the former contains many-body terms. In general, we can write it as
\begin{equation}\label{eq:position_interacting}
    \gamma = \sum_I a_I \Upsilon_I,
\end{equation}
where $I = (j_1s_1, j_2s_2, \dots, j_{|I|}s_{|I|})$ is a multi-index and 
\begin{equation}
    \Upsilon_I = i^{|I|(|I|-1)/2}\prod_{js\in I} \Gamma_j^s
\end{equation}
is a Hermitian $|I|$-body product of single-particle Majoranas. Note that $|I|$ is odd since $\gamma$ is fermionic and therefore contains only terms with an odd number of fermions. 

In a non-interacting system, we can relate the single-particle and ground-state MBSs by 
\begin{equation}\label{eq:sp_to_mb_majorana}
    \gamma = \Gamma Q,
\end{equation}
where $Q$ is the projector on the ground-state sector. It can be expressed in terms of the other Bogoliubov quasiparticles $\chi_n, \tilde\chi_n$ (see \cref{eq:HS_nonint}) as 
\begin{equation}\label{eq:projector}
    Q = \prod_{n=1}^{N_S-1} \frac{(I - i\chi_n \tilde{\chi}_n)}{2}.
\end{equation}
Plugging \cref{eq:projector} into \cref{eq:sp_to_mb_majorana}, we get
\begin{equation}\label{eq:mb_majorana_projected}
\begin{aligned}
    \gamma = \frac{1}{2^{N_S-1}} (\Gamma - i\Gamma \chi_1\tilde{\chi}_1 + \dots) 
    = \frac{\Gamma}{2^{N_S-1}} + M,
\end{aligned}
\end{equation}
where $M$ is an operator containing three-body terms and higher. 
The single-particle content of $\gamma$ is completely determined by $\Gamma$ while the many-body content depends on all the quasiparticles. Note that $\gamma$ has non-zero many-body content, even in the non-interacting case.

The two ground-state MBSs are not independent, as they are related by
\begin{equation}
    \tilde\gamma = iP\gamma,
\end{equation}
where $P = \prod_{j} i\Gamma_j^+\Gamma_j^-$ is the total parity operator. This is different for the single-particle MBSs whose wavefunctions are independent.

Turning to the localization properties of $\Gamma$ and $\gamma$, we first define a bipartition $R,\comp{R}$ of $S$. Then, in the non-interacting system, \cref{eq:non-interacting_pos} can be split into two parts
\begin{equation}
    \Gamma = \sum_{s, j \in R} a_{js} \Gamma_j^s + \sum_{s, j \in \comp{R}} a_{js} \Gamma_j^s,
\end{equation}
where each term lives either in $R$ or $\comp{R}$. The reduced operator is simply
\begin{equation}
    \Gamma_R = d_{\comp{R}}\sum_{s, j \in R} a_{js} \Gamma_j^s.
\end{equation}

In the interacting case, we have to split \cref{eq:position_interacting} into three parts
\begin{equation}\label{eq:locality_interacting}
    \gamma = \sum_{I \in R} a_I \Upsilon_I + \sum_{I \in \comp{R}} a_I \Upsilon_I + 
    \sum_{I \in R\comp{R}} a_I \Upsilon_I,
\end{equation}
where for the sum over $I \in R\comp{R}$, we require that $I$ contains indices in both regions. The last term is a non-local term, which makes the locality of the operator less obvious.

Although $\gamma$ is generally non-local, its effective action within a region $R$ can still be made precise using the partial trace. Under the partial trace to $R$, all terms that are non-trivial in $\comp{R}$ vanish as they are traceless there. Only the first term in \cref{eq:locality_interacting} survives the partial trace, and the reduced operator becomes
\begin{equation}
    \gamma_R = d_{\comp{R}}\sum_{I \in R} a_I \Upsilon_I,
\end{equation}
which captures the part of $\gamma$ that an odd coupling contained in $R$ would effectively probe.

Finally, we comment on another notion of Majorana locality discussed in the literature, namely, commutation properties~\cite{bozkurtInteractioninducedStrongZero2025,obrienManybodyInterpretationMajorana2015}. For a single-particle Majorana operator, if $\Gamma_R = 0$ then it commutes with all even operators $A_R$ in $R$ ($[A_R, \Gamma] = 0$), and anti-commutes with all odd operators $F_R$ in $R$ ($\{F_R,\Gamma\} = 0$). However, due to the last term in \cref{eq:locality_interacting}, this implication does not hold for ground-state MBSs. Therefore, requiring that $\gamma$ (anti-) commutes with all local operators is a stronger condition than $\gamma_R = 0$.

\section{Braiding: Gauge and bounds}\label{app:braiding} 

\begin{figure}[th]
    \centering
    \includegraphics[width=0.9\linewidth]{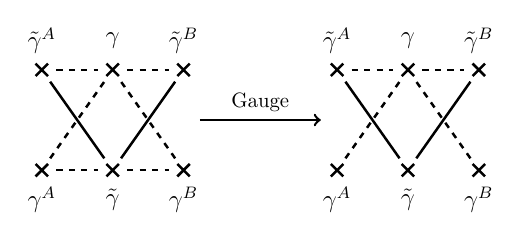}
    \caption{In the gauge that maximizes Majorana separation within each coupling region, the odd part of the coupling Hamiltonian generates four terms for every pair of coupled systems. This is illustrated to the left. To reduce the number of terms, we perform a gauge rotation, leading to the simplified structure to the right.}
    \label{fig:gauge}
\end{figure}
In the gauge described in \cref{sec:gauge}, the odd part of $H_{RR_A}$ and $H_{RR_B}$ will generate eight terms in the effective Hamiltonian, each pairing a ground-state MBS in $S$ with one in $A$ or $B$. To reduce the number of couplings, we transform the gauge according to \cref{fig:gauge}, rotating the MBSs in $A$ and $B$ to eliminate two of the couplings to $S$. If the MBSs are well-separated, only a small gauge rotation is required to remove the couplings. In $S$, we keep the gauge that minimizes $\norm{\gamma_R}_2$. The resulting effective coupling is
\begin{equation}\label{eq:braiding_eff_app}
    \begin{aligned}
       H_{c,\text{eff}} &= gi\gamma\tilde\gamma/2 + \frac{1}{4}\sum_{X=A,B} (2g_X i\gamma^X \tilde\gamma^X
       +  \Delta_X i\tilde\gamma \tilde\gamma^X \\ &+ \tau_X i\gamma \gamma^X +  \tilde\tau_X i\gamma \tilde\gamma^X + U_X \gamma \tilde\gamma \gamma^X \tilde \gamma^X ),
    \end{aligned}
\end{equation}
with coefficients
\begin{subequations}\label{eq:braiding_coeffs}
    \begin{align}
        g &= \sum_{X=A,B} \Tr{ \left[ i\gamma\tilde{\gamma} H_{RR_X}^e\right]}/d_{R_X}, \\
        g_X &= \Tr{\left[ i\gamma^X\tilde{\gamma}^X H_{RR_X}^e\right]}/d_{R}, \\
        \Delta_X &= \Tr{\left[i\tilde{\gamma}\tilde{\gamma}^X H_{RR_X}^o\right]}, \\
        \tau_X &= \Tr{\left[i\gamma\gamma^X H_{RR_X}^o\right]}, \\
        \tilde\tau_X &= \Tr{\left[i\gamma\tilde\gamma^X H_{RR_X}^o\right]}, \\
        U_X &= \Tr{\left[\gamma\tilde{\gamma} \gamma^X\tilde{\gamma}^X H_{RR_X}^e\right]}.
    \end{align}
\end{subequations}

The only two desired terms in $H_\text{eff}$ for the braiding protocol are $\Delta_A$ and $\Delta_B$. A time-dependent term $i\varepsilon \gamma\tilde\gamma$ is also necessary, but we do not include it in $H_c$ (the terms proportional to $g$ and $g_X$ originate from $H_c$). By locality of the ground-state MBSs and their parities, the other coefficients are bounded by
\begin{subequations}\label{eq:braiding_bounds_app}
    \begin{align}
        |g| &\leq \sum_{X=A,B} \norm{(H_{RR_X}^e)_{R}}_p \norm{(i\gamma\tilde\gamma)_{R}}_q /d_{R_X}, \\
        |g_X| &\leq \norm{(H_{RR_X}^e)_{R_X}}_p \norm{(i\gamma^X\tilde\gamma^X)_{R_X}}_q /d_R, \\
        |\tau_X| &\leq \norm{H_{RR_X}^o}_p \norm{\gamma_R}_q \norm{(\gamma^X)_{R_X}}_q, \\
        |\tilde\tau_X| &\leq \norm{H_{RR_X}^o}_p \norm{\gamma_R}_q \norm{(\tilde{\gamma}^X)_{R_X}}_q, \\
        |U_X| &\leq \norm{H_{RR_X}^e}_p \norm{(i\gamma\tilde{\gamma})_R}_q \norm{(i\gamma^X\tilde{\gamma}^X)_{R_X}}_q.
    \end{align}
\end{subequations}
Using the optimized gauge for all systems yields tighter bounds on some of the coefficients but introduces additional couplings.

\section{Bounds on transport}\label{app:transport}
We consider two states $|e\rangle$ and $|o\rangle$ with energy difference much smaller than temperature $T$, and a set of leads indexed by $\alpha$. We will use a rate equation formalism and bound the conductances in the steady state by our quality measures. The transition rates between states is given by Fermi's golden rule as
\begin{align*}
    W_\alpha^{\mathrm{in}}(e\!\to\!o) &= t_\alpha u_\alpha^2 p_\alpha, &
    W_\alpha^{\mathrm{out}}(e\!\to\!o) &= t_\alpha v_\alpha^2 \bar{p}_\alpha, \\
    W_\alpha^{\mathrm{in}}(o\!\to\!e) &= t_\alpha v_\alpha^2 p_\alpha, &
    W_\alpha^{\mathrm{out}}(o\!\to\!e) &= t_\alpha u_\alpha^2 \bar{p}_\alpha,
\end{align*}
where $p_\alpha = p(\mu_\alpha,T)$ is the Fermi-Dirac function, $\bar{p}_\alpha = 1 - p_\alpha$, and $u_\alpha^2 = |\langle o | f_\alpha^\dagger | e \rangle|^2$ and $v_\alpha^2 = |\langle o | f_\alpha | e \rangle|^2$~\cite{bruusManyBodyQuantumTheory2004, tsintzisCreatingDetectingPoor2022}. We assume that $f_\alpha$ removes one fermion from region $R_\alpha$ and is normalized by $\norm{f_\alpha}_p=1$.

We define the total transition rates between the states as
\begin{align}
W_{e\to o} &= \sum_{\alpha}W_{\alpha}^{\mathrm{in}}(e\!\to\!o)+W_{\alpha}^{\mathrm{out}}(e\!\to\!o),\\
W_{o\to e} &= \sum_{\alpha}W_{\alpha}^{\mathrm{in}}(o\!\to\!e)+W_{\alpha}^{\mathrm{out}}(o\!\to\!e),
\end{align}
and $W = W_{e \to o} + W_{o \to e}$. Let $P_e$ and $P_o$ be the probabilities of being in the even and odd state, respectively. Their time evolution is described by the rate equation
\begin{equation}
    \dot{P}_e = -W_{e\to o} P_e + W_{o\to e} P_o,
\end{equation}
and $P_e + P_o = 1$ determines $P_o$. The steady state solution is
\begin{equation}
    P_e = \frac{W_{o\to e}}{W}, \qquad 
    P_o = \frac{W_{e \to o}}{W}.
\end{equation}

To calculate the current in the steady state, it is useful to define the total rate associated with a lead
\begin{align}
    W_{\alpha} &= W_{\alpha}^{\mathrm{in}}(e\!\to\!o) + W_{\alpha}^{\mathrm{out}}(e\!\to\!o) \nonumber \\
    &+ W_{\alpha}^{\mathrm{in}}(o\!\to\!e) + W_{\alpha}^{\mathrm{out}}(o\!\to\!e) = t_\alpha(u_\alpha^2 + v_\alpha^2),
\end{align}
and an asymmetry parameter
\begin{equation}
\eta_{\alpha} = 
u_\alpha^2 - v_\alpha^2.
\end{equation}
The current from lead $\alpha$ in the stationary state is
\begin{align}
    I_\alpha &=P_e \left( W_\alpha^{\mathrm{in}}(e \to o) - W_\alpha^{\mathrm{out}}(e \to o) \right) \nonumber \\
    &+ P_o \left( W_\alpha^{\mathrm{in}}(o \to e) - W_\alpha^{\mathrm{out}}(o \to e) \right)  \nonumber \\
    &= W_\alpha (p_\alpha - 1/2) 
    - \frac{t_\alpha \eta_\alpha}{W} \sum_\beta t_\beta\eta_\beta (p_\beta - 1/2).
\end{align}
The asymmetry parameter $\eta_\alpha$ is related to the Majorana polarization in non-interacting models and can be determined experimentally by measuring the current for different lead voltages or tunnel couplings \cite{dourado2025measuring}. In interacting models, we have the bound 
\begin{equation}\label{eq:delta_bound}
    |\eta_\alpha| \leq \norm{\gamma_{R_\alpha}} _q\norm{\tilde{\gamma}_{R_\alpha}}_q,
\end{equation}
where $q = 1-1/p$, see derivation below.

A direct experimental demonstration of overlapping MBSs is to measure the conductance 
\begin{align}
        G_{\alpha\beta} &= \frac{\partial I_\alpha}{\partial \mu_\beta} = p'_\beta\left(
        W_\alpha \delta_{\alpha\beta} -\frac{t_\alpha t_\beta\eta_\alpha \eta_\beta}{W}\right),
\end{align}
where $p'_\beta = \partial p_\beta / \partial \mu_\beta$.  The non-local conductance ($\alpha \neq \beta$) is finite only if both $\eta_\alpha \neq 0$ and $\eta_\beta \neq 0$, which implies that there are overlapping MBSs on both leads.

\subsubsection{Bounding $\eta_\alpha$}
We want to derive the bound \cref{eq:delta_bound}. The projection of $f_\alpha^\dagger$ onto the ground-state manifold can be written in terms of the ground-state MBSs as
\begin{equation}
    Qf_\alpha^\dagger Q = a \gamma/2 +b \tilde{\gamma}/2,
\end{equation}
where $a = \Tr[\gamma f_\alpha^\dagger]$ and $b = \Tr[\tilde{\gamma} f_\alpha^\dagger]$. These are bounded as $|a| \leq \norm{\gamma_{R_\alpha}}_q \norm{f_\alpha^\dagger}_p = \norm{\gamma_{R_\alpha}}_q$ and $|b|\leq \norm{\tilde{\gamma}_{R_\alpha}}_q$. Then we have 
\begin{align}
    |\eta_\alpha| &= \left| |\langle e|f_\alpha^\dagger|o\rangle|^2-|\langle{e}|f_\alpha|{o}\rangle|^2 \right|\nonumber \\
    &= \left||a+ib|^2-|\bar{a}+i\bar{b}|^2\right|/4 \nonumber \\
    &= \left|\mathrm{Im}\!\left[a\bar{b}\right]\right| \leq \norm{\gamma_{R_\alpha}}_q \norm{\tilde{\gamma}_{R_\alpha}}_q.
\end{align}

\subsubsection{Summing over leads}
If we sum the current from a set of leads $A$ and vary the chemical potential of a disjoint set of leads $B$ (for example, corresponding to different spins), we get
\begin{equation}
    \sum_{\substack{\alpha\in A \\ \beta \in B}} G_{\alpha\beta} = -\frac{1}{W}\sum_{\alpha\in A} t_\alpha \eta_\alpha \sum_{\beta \in B} p'_\beta t_\beta \eta_\beta.
\end{equation}
We will use $\norm{\tilde{\gamma}_R}\leq 2$, and assume that all leads in $B$ have the same temperature and chemical potential. Then we have
\begin{equation}
    \left|\sum_{\alpha \in A} t_\alpha \eta_\alpha\right| \leq  t_A Q_o^A ,
\end{equation}
where
\begin{subequations}
\begin{align}
    Q_o^A =& \sqrt{\sum_{\alpha \in A} \norm{\gamma_{R_\alpha}}^2_q}, \\
    t_A =& \sqrt{\sum_{\alpha \in A} |t_\alpha|^2}.
\end{align}
\end{subequations}
The non-local conductance then satisfies 
\begin{equation}
    \Bigl|\sum_{\substack{\alpha\in A \\ \beta \in B}} G_{\alpha\beta} \Bigr| \leq \frac{p'_B}{W} t_A t_B Q_o^A Q_o^B. 
\end{equation}
This is the form we use in \cref{sec:example_abs_to_ysr}.

\end{document}